\documentclass[a4paper,11pt]{article}
\pdfoutput=1 

\usepackage{jheppub} 

\usepackage{subfigure}
\usepackage[T1]{fontenc} 

\usepackage{slashed}			
 
\title{\boldmath Light Mediators in Anomaly Free $U(1)_X$ Models I - Theoretical Framework}

\author[a,1]{F.C. Correia,}
\author[b,c,2]{Svjetlana Fajfer}

\affiliation[a]{Institut f\" ur Physik, Technische Universit\" at Dortmund, D-44221 Dortmund, Germany}
\affiliation[b]{Department of Physics, University of Ljubljana, Jadranska 19, 1000 Ljubljana, Slovenia}
\affiliation[c]{J. Stefan Institute, Jamova 39, P. O. Box 3000, 1001 Ljubljana, Slovenia}

\emailAdd{fagner.correia@tu-dortmund.de}
\emailAdd{svjetlana.fajfer@ijs.si}

\abstract{
We examine  theoretical features of $U(1)_X$ extensions of the Standard Model whose quantum anomalies are canceled per generation. Similarly to other versions, the theory consists of a Two-Higgs-Doublet Model plus a scalar singlet embedded into the $SM \otimes U(1)_X$ gauge group, and introduces small modifications to the $Z$-boson interactions. These changes can be minimized by exclusively charging right-handed fermions under the new Abelian symmetry, and are compensated by the neutral $X$-boson exchange. Non-universality of fermion couplings can also be achieved by requiring one single $X$-charged family.  
In general,  $X$ gauge bosons can be separated into $A'$ (dark photons) and $Z'$ subsets, distinguished by the presence of axial-vector currents. 
$A'$ physics is commonly simpler to constrain and therefore favored by experimental tests. Finally, the model can be UV completed both by stable $\chi$ fermions or by right-handed  neutrinos. The prior case may provide cold WIMPs in the theory. 
}

\begin{document} 
\maketitle
\flushbottom

\section{Introduction}
\label{sec:intro}
The  existence  of dark matter (DM) has  been supported by  numerous astrophysical and cosmological arguments (see e.g. \cite{Peter:2012rz}). 
On the other hand,  experimental and theoretical efforts for direct detection have  provided strong constraints on the couplings of DM candidates to the Standard Model (SM) particles.

The simplest DM model can be obtained by extending the SM with an additional $U(1)_X$ gauge symmetry. This type of extension commonly requires, for the UV completion, at least one new fermion $\chi$ which can be made stable by some ad hoc dark symmetry. The symmetry forbids the appearance of tree-level couplings between $\chi$ and SM fields  and a  DM portal is generated exclusively through the neutral gauge bosons \cite{Correia:2016xcs}. 
In the case where the boson associated with the new Abelian symmetry contains only vector couplings it is generally referred to as a \textit{dark photon}.  

Very often dark photons searches are performed by assuming that the new vector is on-shell and then decays into $\mu^+\mu^-$ pairs \cite{TheBABAR:2016rlg}. 
New Physics effects in the few MeV region generally require searches for $e^+ e^-$ channels facing a vast set of background events. The possible discrepancies \cite{Feng:2016ysn,TuckerSmith:2010ra} in the region $m_X < 2m_\mu$ motivate proposals of the simplest  $U(1)_X$ SM extension containing one weakly coupled gauge boson.  As summarized by the authors of Ref. \cite{Batell:2011qq} this task must follow in accordance with some strong requirements, which goes from the absence of new electrically charged fields at low energies to the consistency of neutrino interactions with electrons and nucleons. In particular, 
the last argument will motivate the proposal of right-handed specific models.

The present work contains a detailed study of the basic properties of $U(1)_X$ models. We are guided by general principles as (i) introduction of a minimal set of free parameters and physical degrees of freedom (falsifiability); (ii) the fermions are accommodated in the SM representations  
and quantum anomalies are canceled per generation (as in the SM); (iii) the degree of fermion non-universality should explain 
observed  discrepancies between theory predictions and experimental measurements, such as the flavor anomalies in B meson decays \cite{Hiller:2003js,Hiller:2017bzc} and the muon anomalous magnetic moment puzzle \cite{Bennett:2006fi,Bennett:2004pv,Jegerlehner:2009ry,Hagiwara:2011af,Davier:2010nc}. In a subsequent work we consider manifestation of 
New Physics in the low energy regime ($\sim$ MeV) by allowing small couplings of the new vector $X$ to matter fields of the order  $g_X \sim 10^{-4}$ to $10^{-2}$. In Ref. \cite{Ilten:2018crw} one can notice, for instance, that the region 10 MeV - 1 GeV still allows the presence of $X$ with a coupling $g_X \approx 10^{-4}$ in the $U(1)_{B-L}$ case, whose strength might imply measurable effects in heavy meson physics. The same is not true in the case of $U(1)_B$ or the protophobic model. This distinction illustrates the impact of different $X$-hypercharge assignments albeit under the same gauge structure. 

A minimal SM extension is readily achieved even when non-universal $X$-hypercharges for leptons and quarks are considered. The first consequence of arbitrary charges is  the appearance of a flavor matrix $\mathbb{F}$ in the fermion-gauge sector. Its  properties are specially interesting in the case when only one generation is charged under $X$. Moreover, the attempt to reduce the number of free parameters may favor the choice of a chiral $U(1)_X$. 
Once more, if we aim to control the modification of neutrino interactions, the right-handed (RH) currents might be the appropriate option within this  framework. 
In the scenario where only RH fermions are charged under $U(1)_X$, the constraints from quantum anomalies arise exclusively from $U(1)^3_X$, $ U(1)^2_X U(1)_Y$, $U(1)_X U(1)^2_Y$ and $SU(3)^2 U(1)_X$ currents, once $SU(2)$ does not play any role. This minimal version\footnote{Here, the term `minimal' is related to  the anomaly cancellations.} shows that is not possible to charge the quarks universally in order to avoid flavor changing neutral currents (FCNC) and to preserve, at the same time, the feature of anomaly cancellations per generation. Besides, the call for fermion non-universality along with the criterion to recover a consistent CKM matrix will demand the introduction of a new scalar doublet.

In summary, here we introduce a Two-Higgs-Doublet Model (2HDM) embedded in the $SM\otimes U(1)_X$ gauge group where only the second generation of fermions is charged under $X$. Our main goal is to determine the allowed region of the parameter space in comparison with similar $U(1)$ versions already presented, for instance, in \cite{Feng:2016ysn} and \cite{Babu:2017olk},  whose phenomenology were focused in the MeV-GeV regime.  Motivated by   the absence of left-handed (LH) singlets under the SM, 
we construct the UV completion by including in the setup a new right-handed $\chi_R$ fermion charged exclusively under $U(1)_X$. We focus on a stable fermion whose mass is generated after a symmetry breaking performed by a scalar singlet.  
The alternative version where a leptonic Yukawa Lagrangian generates neutrino masses is possible too and is  discussed in the  Appendix \ref{Ap.FermionChi}. Notwithstanding, we choose to investigate in detail a DM model due to the following reason - the tiny couplings governing the $\chi \chi \leftrightarrow SM$ portal can turn a Weakly Interacting Massive Particle (WIMP) overabundant in the present universe and therefore forbidden. Hence, the relic abundance constraints are expected to cover the parameter space in the opposite direction if compared with those from direct detection, which provide us with a mechanism to maximize its excluded area and eventually rule out the particular model. 

The work is divided into three sections. Section \ref{Sec.II} introduces the general features of $U(1)_X$ theories, and motivates the variant with $X$-charged right-handed fermions. Section \ref{Sec:DPandZ} discusses the experimental status for dark photons, $A'$, and $Z'$ searches. The last section is devoted to our conclusions.

\section{General properties of \texorpdfstring{$U(1)_X$}{} and inclusion of right-handed fermions}\label{Sec.II}

In this section we determine the functional dependence of the vector and axial-vector currents on the hypercharges and fundamental constants, like v.e.v's and couplings. We see that general Abelian extensions must produce $Z$-like, instead of $A$-like, vertexes, i.e.
\begin{equation}\label{Eq.axial}
\mathcal{L} \ \supset \ \frac{x_V^f}{2} \bar{f} \gamma_\mu f X^\mu + \mathcal{L}_\chi \quad \rightarrow \quad \frac{1}{2} \bar{f} \gamma_\mu (x_V^f + x_A^f \gamma_5) f X^\mu + \mathcal{L}_\chi\, , 
\end{equation} 
where $x_V, x_A$ are non-universal and flavor violating matrices and $\mathcal{L}_\chi$ describes the possible interactions of new fermions in the theory. The Eq.(\ref{Eq.axial}) illustrates the fact that the dark photons phenomenology comprises a subset of the $X$ boson theory, where the axial couplings are set to zero. 
By keeping the complete vertex in our study,  we can determine the impact of $x_A$ on the allowed parameter space. In addition, non-universality effects accommodated within the $SM \otimes U(1)_X$ model usually spread into both quark and leptonic sectors. Apart from that, the long-lived boson is commonly assumed to decay exclusively into a dark-sector, i.e. $Br(X \rightarrow \chi \chi) = 1$, which motivates experimental searches in processes  with invisible final states. By allowing the gauge boson to couple to  an electron-positron pair, the parameter space for $X$ (long-lived enough to decay out of the detector) has the impact of loosening important bounds, such as those from $K \rightarrow \mu + \text{invisibles}$ \cite{Pang:1989ut}. 

Apart from the Yukawa Lagrangian and the X-hypercharge assignments, the following description and results are general in the sort of 2HDM plus scalar singlet of $SM \otimes U(1)_X$:
\begin{itemize}
	\item Three vector fields $W^\mu$ from $SU(2)_L$, one vector $B_\mu^Y$ from $U(1)_Y$ and $B_\mu^X$ from $U(1)_X$;
	\item Three independent coupling constants $g, g_Y, g_X$  and a kinetic mixing constant $\epsilon$;
	\item Three generations of weak isospin doublets:
	\begin{equation}
	(L_L)_i = \begin{pmatrix}
	\nu_i \\ e_i
	\end{pmatrix}_L \, ,\qquad 
	(Q_L)_i = \begin{pmatrix}
	u_i \\ d_i
	\end{pmatrix}_L \,,
	\end{equation}
	with $i = 1,2,3$;
	\item right-handed $SU(2)_L$ singlets: 
	$\chi_{R}, l_{i R}, u_{i R}, d_{i R}$;
	\item $Y$ hypercharges: \begin{equation}
	Y_L = -\frac{1}{2}; \quad Y_Q = \frac{1}{6}; \quad Y_l = -1; \quad Y_\chi = 0; \quad Y_u = \frac{2}{3}; \quad Y_d = - \frac{1}{3}\,;
	\end{equation}
    \item $X$ hypercharges: 
	\begin{equation}
	X_L = 0; \quad X_Q = 0; \quad X_{e2} = 1; \quad X_{\chi_R} = -1; \quad X_{u2} = -1; \quad X_{d2} = 1\,;
	\end{equation}
	with the remaining RH fields uncharged.
	\item Higgs doublets $\phi^0$, $\phi^X$ and singlet $s$: 
	\begin{equation}\label{Eq.ScHyp}
	Y_{0} = Y_{X} = \frac{1}{2}; \quad X_{0} = 0; \quad X_{X} = -1; \quad 
	Y_s = 0 \quad X_s = 1\,.
	\end{equation}
	Notice that the term $\phi_0^{\dagger} \phi_X s$ is allowed in the scalar potential.
	
	\item Electroweak Lagrangian is
	\begin{subequations}\label{Eq.EleLag1}
		\begin{eqnarray}
		\mathcal{L} &=& - \frac{1}{4} \mathbf{W}^{\mu \nu} \cdot
		\mathbf{W}_{\mu \nu} 
		- \frac{1}{4} B^{Y \mu \nu} B^Y_{\mu \nu}
		- \frac{1}{4} B^{X \mu \nu} B^X_{\mu \nu} 
		+ \frac{\epsilon}{2} 
		B^{Y \mu \nu} B^X_{\mu \nu} +
	\label{Eq.kinmix}	\\ 
		& &  + 	(D_\mu \phi^0)^\dagger (D^\mu \phi^0)
		+
		(D_\mu \phi^X)^\dagger (D^\mu \phi^X) 
		+ 
		(D_\mu s)^\dagger (D^\mu s)
		-
		V(\phi^0,\phi^X,s)
		\\ 
		& & - \sum_{\alpha = 1,2,3} \left(\sum_{\beta = 1,3} \overline{L}_{\alpha L} \phi^0 Y^l_{\alpha \beta} e_{\beta R}
		 + \overline{L}_{\alpha L} \phi^X Y^l_{\alpha 2} e_{2 R} + h.c.\right)
		\\ 
		& & - \sum_{\alpha = 1,2,3} \sum_{\beta = 1,3} \left(\overline{Q}_{\alpha L} \phi^0 Y^D_{\alpha \beta} d_{\beta R} + \overline{Q}_{\alpha L} \tilde{\phi}^0 Y^U_{\alpha \beta} u_{\beta R} + h.c.\right)
		\\ 
		& & - \sum_{\alpha = 1,2,3} \left(\overline{Q}_{\alpha L} \phi^X Y^D_{\alpha 2} d_{2 R} + \overline{Q}_{\alpha L} \tilde{\phi}^X Y^U_{\alpha 2} u_{2 R} + h.c.\right)
		\\ 
		& &  - Y_s \ \overline{\chi_{L}} \chi_{R} s - Y^*_s \ \overline{\chi_{R}} \chi_{L} s^* + \label{Eq.ChiYuk}
		\\ 
		& & + i \sum_{\alpha = 1,2,3} \biggl[ \overline{L}_{\alpha L} \slashed{D} {L}_{\alpha L} 
		+
		\overline{Q}_{\alpha L} \slashed{D} {Q}_{\alpha L} 
		+ \\ 
		& & + \overline{l}_{\alpha R} \slashed{D} {l}_{\alpha R} 
		+ 
		\overline{d}_{\alpha R} \slashed{D} {d}_{\alpha R} 
		+ 
		\overline{u}_{\alpha R} \slashed{D} {u}_{\alpha R} 	\biggr]
		+ i 
		\overline{\chi}_{R} \slashed{D} {\chi}_{R} 	\,. 
		\end{eqnarray}
	\end{subequations}
\end{itemize}	
\paragraph{Anomalies}

A basic prerequisite of any ultraviolet complete gauge theory is that it is free of  triangle anomalies  \cite{Ellis:2017nrp}. 
The following equations summarize how this criterion can be achieved for arbitrary $X$ charges within  $SM\otimes U(1)_X$ models. 
Again, $Q$ and $L$ denote quark and lepton doublets, respectively, while the rest denoted by $u_R$, $d_R$, $e_R$ refer to the charges of right-handed fields:
	\begin{subequations}\scriptsize	\begin{align}
	X^3 &: 2 \left( \sum_{L} X_L^3 + 3 \sum_{Q} X_Q^3 \right) 
	- \left[\sum_{l_R} X_{l_R}^3 + \sum_{\chi_R} X_{\chi_R}^3 + 
	3 \left(\sum_{U_R} X_U^3 + \sum_{D_R} X_D^3\right)\right] = 0\,, \label{eq.ANX3}	
	\\
	Y X^2 &:
	2 \left( \sum_{L} Y_L X_L^2 + 3 \sum_{Q} Y_Q X_Q^2 \right) 
	- \left[\sum_{l_R} Y_{l_R} X_{l_R}^2 + \sum_{\chi_R} Y_{\chi_R} X_{\chi_R}^2 + 
	3 \left(\sum_{U_R} Y_U X_U^2 + \sum_{D_R} Y_D X_D^2\right)\right] = 0 \,,\label{eq.ANYX2}
	\\
	Y^2 X &: \text{Same as Eq.(\ref{eq.ANYX2}) with $X \leftrightarrow Y$}\,,
    \\
    SU(2)^2 X &:
	\left(\sum_{L} X_L + 3 \sum_{Q} X_Q \right) = 0\,,
	\\
	SU(3)^2 X &:
	\left(2 \sum_{Q} X_Q - \sum_{U_R} X_U - \sum_{D_R} X_D\right) = 0\,,
	\\
	grav^2 X &:	
	2 \left( \sum_{L} X_L + 3 \sum_{Q} X_Q \right) 
	- \left[\sum_{l_R} X_{l_R} + \sum_{\chi_R} X_{\chi_R} + 
	3 \left(\sum_{U_R} X_U + \sum_{D_R} X_D \right)\right] = 0 \,. \label{eq.ANgrav}
	\end{align}
	\end{subequations}

In the Standard Model the above equations are solved per generation. This property can be taken as part of the SM structure and implies that no information about a new fermion family could be found particularly through these diagrams. Here we will follow this principle. The solutions define a subset of $U(1)$ extensions and are given by\footnote{The solutions are obtained by solving to $X_L$  in the quadratic expressions, with $Y_\chi = 0$, and then for the gravitational and cubic case.}\cite{DelleRose:2017xil} 
\begin{equation}\label{Eq.an}
X_D = 2 X_Q - X_U, \quad X_L = -3 X_Q, \quad X_l = -2X_Q - X_U, \quad X_\chi = X_U - 4X_Q\, .
\end{equation}

\paragraph{Two Higgs Doublets Requirement} The theory is designed to be (a) non-universal under $X$ charges and (b)  to be minimal in its particle content. Thus, let us first assume it is possible to generate all fermion masses through only one Higgs doublet. FCNC should not appear in the scalar sector under this property. In order to construct the Yukawa Lagrangian, one has to consider solutions to the equations 
\begin{equation}
X_{L_i} - X_{l_i} - X_0 = 0; \quad 
X_{Q_i} - X_{U_i} + X_0 = 0; \quad
X_{Q_i} - X_{D_i} - X_0 = 0\, ,
\end{equation} 
which should also satisfy Eq.(\ref{Eq.an}). In addition, the condition
\begin{equation}
X_{L_i} - X_{l_i} = X_{L_j} - X_{l_j}\,,
\end{equation}
is necessary for generating masses for all charged fermions\footnote{Alternatively, one could create the entries $X_{L_i} - X_{l_j} = X_{L_j} - X_{l_i}$ along the diagonal ones $X_{L_i} - X_{l_i} = X_0$, which would lead to the same conclusion.}, where the indexes denote the two non-universal X-hypercharges. Notwithstanding, the mass matrix corresponding to the above criterion can be put in a block-diagonal form, and would be inconsistent with the expected form of the mixing matrices. We must additionally fill at least two more entries by requesting the condition
\begin{equation}
X_{L_i} - X_{l_j} = X_{L_j} - X_{l_j}\,.
\end{equation}
Now, both conditions combined imply 
\begin{equation}\label{Eq.NonUni}
X_{L_i} = X_{L_j},  \qquad X_{l_i} = X_{l_j}\,,
\end{equation}
hence breaking the (a) criterion. Naturally, in the $U(1)_X$ extensions one must introduce a larger scalar sector, in comparison to the SM, due to the creation of the longitudinal polarization for the  massive $X_\mu$ boson. In summary, the Eq.(\ref{Eq.NonUni}) will state that any version containing only one Higgs doublet is necessarily universal in the fermion families.  

\subsection{Kinetic mixing}
Once $\epsilon$ is assumed to be  a small parameter, it is convenient to translate its dependence directly into the coupling constants, thus leaving the kinetic Lagrangian in a diagonal form. The task can be achieved  through the field redefinition
\begin{equation}\label{Eq:Red}
B^Y_{\mu} \rightarrow B^Y_{\mu} + \epsilon B^X_{\mu}\, ,
\end{equation}
i.e., by rewriting Eq.(\ref{Eq.kinmix})
\begin{equation}
\mathcal{L}_{k.m.} \supset
- \frac{1}{2} (B^Y_{\mu} + \epsilon B^X_{\mu})  \hat{\mathcal{O}}^{\mu \nu}  
(B^Y_{\nu} + \epsilon B^X_{\nu})
- \frac{1}{2} B^X_{\mu}  \hat{\mathcal{O}}^{\mu \nu}  
B^X_{\nu} 
+ \epsilon \
B^{X}_\mu \hat{\mathcal{O}}^{\mu \nu}  
(B^Y_{\nu} + \epsilon B^X_{\nu})\,,
\end{equation}
where $\hat{\mathcal{O}}^{\mu \nu} = \partial^\mu \partial^\nu - \partial^2 g^{\mu \nu}$. Up to order $\mathcal{O}(\epsilon)$,
\begin{equation}
\mathcal{L}_{k.m.} \supset
- \frac{1}{2} B^Y_{\mu}  \hat{\mathcal{O}}^{\mu \nu}
B^Y_{\nu}
- \frac{1}{2} B^X_{\mu}  \hat{\mathcal{O}}^{\mu \nu}
B^X_{\nu} + \mathcal{O}(\epsilon^2)\,,
\end{equation}
i.e. the crossed terms vanishes and the mixing effect is converted into a new term in the covariant derivative:
\begin{equation}
D_\mu \rightarrow D_\mu = \partial_\mu -i g \mathbf{W}_\mu \cdot \mathbf{\tau} - ig_Y B_\mu^Y Y^p - i(\kappa Y^p +  g_X X^p) B_\mu^X \,,
\end{equation}
where, up to first order, one may write $\epsilon g_Y \equiv \kappa$. 

\subsection{Couplings and masses of gauge bosons}
In the previous section we showed that the non-universal model must contain at least two Higgs doublets, here denoted as $\phi^0$ and $\phi^X$, as a necessary condition to recover the correct mass spectrum of the fermions. In addition, a singlet $s$ is required to couple to the fermion $\chi_R$ (or to generate the mixing between the second and the remaining generations of RH neutrinos, see \ref{Ap.FermionChi}) as well as to break a residual $U(1)$ in the potential which could leave the theory with a massless pseudo-Goldstone boson at tree-level \cite{Babu:2017olk}.

The gauge boson masses are extracted from the kinetic piece of the scalar Lagrangian once the scalars acquire a vacuum expectation value. In terms of ladder operators the covariant derivatives can be written as\footnote{Note that, in the basis of Eq.(\ref{Eq:Red}) and since $Y_s = 0$, the kinetic term does not enter in the interactions with the singlet scalar.}
\begin{align}\label{Eq.NeutralGB}
D_\mu \phi^p &= \left[\partial_\mu - i\frac{g}{\sqrt{2}}(W^+ \mathbb{I}_+ + W^- \mathbb{I}_-) -ig \tau_3 W_\mu^3  -ig_Y Y^p B_\mu^Y 
-i (\kappa Y^p +  g_X X^p) B_\mu^X \right]\phi^p \,, \\
D_\mu s &= (\partial_\mu -ig_X X^s B_\mu^X) s \,.
\end{align}  

Once the relation $Q = T_3 + Y$ to the electric charge matrix is preserved, it follows that
\begin{equation}
\phi_0 = 
\begin{pmatrix}
\varphi_0^+ \\ \frac{v_0 + H_0 + i \chi_0}{\sqrt{2}}
\end{pmatrix}, \qquad
\phi_X = 
\begin{pmatrix}
\varphi_X^+ \\ \frac{v_X + H_X + i \chi_X}{\sqrt{2}}
\end{pmatrix}, \qquad
s = \frac{v_s + H_s + i \chi_s}{\sqrt{2}}.
\end{equation}
The charged currents are untouched and result for the $W$ mass $m_W = \frac{g v}{2}$, with $v^2 \equiv (v_0^2 + v_X^2)$. The neutral fields must mix and their mass matrix is extracted from the symmetric expression $\sum_{i = 1}^3 (a_i W^3_\mu+ b_i B^Y _\mu+ c_i B^X_\mu)^2$, whose determinant is zero (massless photon). In the $(W^3_\mu, B^Y_\mu, B^X_\mu)$ basis it is given by
\begin{equation}\label{Eq.MassNGB}
\mathbb{M}^0 = \frac{v^2}{8} 
\begin{pmatrix}
g^2 & -g g_Y &  g (2g_X c_\beta^2 - \kappa) \\
-g g_Y & g_Y^2 & -  g_Y (2g_X c_\beta^2 - \kappa)  \\
g (2g_X c_\beta^2 - \kappa) & -  g_Y (2g_X c_\beta^2 - \kappa) & 4 [g_X^2 \frac{\bar{v}^2}{v^2} - g_X \kappa c_\beta^2]  + \kappa^2
\end{pmatrix}\,,
\end{equation} 
where \begin{equation}
v^2 \equiv (v_0^2 + v_X^2), \qquad \bar{v}^2 \equiv (v_s^2 + v_X^2), \qquad 
c_\beta^2 = \frac{v_X^2}{v^2}. \label{vevs}
\end{equation}
The above real symmetric matrix eigenvectors define an orthonormal basis and compose the orthogonal matrix $\mathbb{V}$ which rotates the fields from the gauge  to the mass basis. Although the choice of parametrization for $\mathbb{V}$ is not physical, there are options which can make the analysis simpler. Consider, for example, the choice made in terms of the three Euler angles in the usual $z x z$ rotations by the angles $(\phi, \theta, \psi)$ (using notation $sin \alpha \equiv s_\alpha$ and $cos\alpha \equiv c_\alpha$)
\begin{equation}
\mathbb{V} = 
\begin{pmatrix}
c_\psi c_\phi - c_\theta s_\phi s_\psi &
c_\psi s_\phi + c_\theta c_\phi s_\psi &
s_\theta s_\psi
\\
-s_\psi c_\phi - c_\theta s_\phi c_\psi &
- s_\psi s_\phi + c_\theta c_\phi c_\psi &
c_\psi s_\theta 
\\
s_\theta s_\phi &
- s_\theta c_\phi &
c_\theta
\end{pmatrix}.
\label{Vmatrix}
\end{equation}  
The angle  $\theta$  introduces the  mixing of $B^X_\mu$ with the remaining gauge fields,  i.e. $\theta$ is the angle between the  $B^X_\mu$ and the z-plane where the $W^3_\mu - B^Y_\mu$ mixing occurs. All three angles can be written using  the couplings, vev's and scalar charges.  


One can notice from Eq.(\ref{Eq.MassNGB}) that the block $(W^3_\mu, B^Y_\mu)$ has a null determinant. This substructure of $\mathbb{M}^0$ implies a zero entry in one eigenvector, which fixes one of the angles. Therefore, by taking $c_\psi = 0; s_\psi = -1$ and applying a phase redefinition $s_\phi \leftrightarrow c_\phi$, $c_\theta \leftrightarrow s_\theta$ we can parametrize $\mathbb{V}$ as
\begin{equation}\label{Eq.param}
\mathbb{V} = 
\begin{pmatrix}
s_\theta c_\phi &
- s_\theta s_\phi &
- c_\theta
\\
s_\phi &
c_\phi &
0 
\\
c_\theta c_\phi &
- c_\theta s_\phi &
s_\theta
\end{pmatrix}.
\end{equation}      
The minimal coupling in the covariant derivative can be presented by
\begin{eqnarray}
	\langle \mathbf{g}^0 | \mathbf{B}_\mu^0 \rangle \rightarrow 
	\langle \mathbf{g}^0 | \mathbb{V}^\intercal \mathbb{V} |\mathbf{B}_\mu^0 \rangle
	&=& \langle \mathbb{V} \mathbf{g}^0 | \mathbb{V} \mathbf{B}_\mu^0 \rangle 
	=\langle \mathbf{g} | \mathbf{B}_\mu \rangle,
\end{eqnarray} 
where the vectors are defined as 
\begin{equation}
\mathbf{B}_\mu \equiv (X_\mu, A_\mu, Z_\mu)^\intercal = \mathbb{V} \mathbf{B}_\mu^0, \qquad  \qquad \mathbf{g} \equiv  (g_R, e \mathbb{Q}, g_Z)^\intercal = \mathbb{V} \mathbf{g}^0
\end{equation}
and \begin{equation}
\mathbf{g}^0 \equiv (g\tau^3, g_Y Y, g_X X + \kappa Y)^\intercal \quad , \quad \mathbf{B}_\mu^0 \equiv (W_\mu^3, B_\mu^Y, B_\mu^X)^\intercal \,.
\end{equation} 
The parametrization of Eq.(\ref{Eq.param}) corresponds to 
\begin{equation}
e \mathbb{Q} =  g s_\phi \tau^3 + g_Y c_\phi Y.
\end{equation} 
By taking $g, g_Y$ to be the same as in the SM,  the angle $\phi$  must correspond to the weak mixing angle, related to  the electric charge by
\begin{equation}
g s_\phi = g_Y c_\phi = e. 
\end{equation}
On the other hand, the Z-couplings will be replaced by
\begin{eqnarray}\label{Eq.gZ}
g_Z &=& c_\theta (g c_\phi \tau_3 - g_Y s_\phi Y) + s_\theta (\kappa Y + g_X X)
= c_\theta g_Z^{SM} + s_\theta (\kappa Y + g_X X), 
\end{eqnarray}
where $g_Z^{SM} = \frac{g}{c_\phi} (\tau_3 -s^2_\phi \mathbb{Q})$, thus making explicit how $\theta$ (i.e the small parameter $s_\theta$) tunes the change in the Z interactions due to the presence of a new neutral gauge boson. Finally, the interactions with the new $X_\mu$ is governed by 
\begin{eqnarray}\label{Eq.gR}
g_R &=& s_\theta (g c_\phi \tau_3 - g_Y s_\phi Y) - c_\theta (\kappa Y + g_X X) 
= s_\theta g_Z^{SM} - c_\theta (\kappa Y + g_X X).
\end{eqnarray}

Once the parametrization of Eq.(\ref{Eq.param}) is established, the above results are general for 2HDM-like models. The fermion interactions with the new $X^\mu$ gauge boson can be determined using Eq.(\ref{Eq.gR}). It contains a term proportional to the SM Z-coupling, weighted by $s_\theta$, and its last piece will regulate the amount of flavor violation in the gauge sector, along with the  gauge symmetry-mass basis  rotation matrices. One can notice that, after electroweak symmetry breaking (EWSB), the chiral version with RH fields still generates the gauge boson vertexes with LH currents via the kinetic coupling $\kappa$. The gauge interactions with X-neutral fields are further suppressed by $s_\theta$. 

Using the parametrization Eq.(\ref{Eq.param}) and the choice of Eq.(\ref{Eq.ScHyp}), $ s_\theta$  can be written as
\begin{equation}
 s_\theta= [1 + (\bar{g}\lambda)^2]^{-\frac{1}{2}}\,,
\end{equation}
where
\begin{equation}
\label{eq:x}
\begin{split}
\lambda &=  \frac{\bar{g}^2  - a_1 + [\bar{g}^2 (\bar{g}^2 + 4 a_2^2) - 2 \bar{g}^2 a_1 + a_1^2]^\frac{1}{2}}{2 \bar{g}^2 a_2}\,,
\\
\bar{g}^2 = g_Y^2 + g^2  \,, \qquad
a_1 &= 4 \left[ g_X^2 \frac{\bar{v}^2}{v^2} - g_X \kappa c^2_\beta \right] + \kappa^2 \,, \qquad
a_2 = 2 g_X c^2_\beta - \kappa \,.
\end{split}
\end{equation}
From these definitions the neutral gauge boson masses, in terms of couplings and v.e.v's, are given by
\begin{eqnarray}
m^2_Z &=& \frac{v^2}{4} \frac{1}{2} \left[\bar{g}^2 + a_1 + [\bar{g}^2 (\bar{g}^2 + 4 a_2^2) - 2 \bar{g}^2 a_1 + a_1^2]^\frac{1}{2}\right] ,\label{Eq.Zmass}\\
m^2_X &=& \frac{v^2}{4} \frac{1}{2} \left[\bar{g}^2 + a_1 - [\bar{g}^2 (\bar{g}^2 + 4 a_2^2) - 2 \bar{g}^2 a_1 + a_1^2]^\frac{1}{2}\right]. \label{Eq.Xmass}
\end{eqnarray}
Now one can verify that, if $g_X, \kappa \ll \bar{g}$
\begin{eqnarray}\label{Eq.limitMass}
m^2_Z \rightarrow \frac{v^2}{4} \bar{g}^2,  \qquad 
m^2_X \rightarrow \frac{v^2}{4} a_1.
\end{eqnarray}
Again, since $\frac{\bar{v}^2}{v^2} \gg c^2_\beta$, it follows that $m^2_X \rightarrow g_X^2 \bar{v}^2 + \kappa^2 \frac{v^2}{4}$. In the same limit the angle may be written in a simplified expression:
\begin{equation}\label{Eq.steta}
s_\theta \approx \frac{|2 g_X c^2_\beta - \kappa|}{\bar{g}} \left[1 - \frac{m^2_X}{m^2_Z}\right]^{-1}.
\end{equation} 
As mentioned before, the above description is general in 2HDM embedded in a $U(1)_X$ extension of the SM. The main feature of this study is: when the charges under $X$ are non-universal, there must appear flavor changing neutral currents both in the interactions with $Z_\mu$ and $X_\mu$, doubly suppressed in the first case by the factor $s_\theta g_X$. Moreover, regarding  the lepton flavor violation (LFV), the first effect of charging LH fields under $X$ 
is that a number of free parameters appear, due to the 
 matrix elements of  $V^l_L$ which rotates these fields to their mass eigenstates. Therefore, the choice for charging only one chiral fermion provides a minimal description of flavor violating processes, although it certainly does not resolve the entire combination of operators. For instance, in the case of $R_{K^{(*)}}$ anomalies in B physics,  the $b \to s \mu^+ \mu^-$ transition  \cite{Hiller:2017bzc} get contributions from the operators $O_9\sim(\bar s \gamma_\mu (1-\gamma_5) b)  (\bar l \gamma^\mu l)$ and $O_{10}\sim(\bar s \gamma_\mu (1-\gamma_5) b)  (\bar l \gamma^\mu \gamma_5 l) $, which can be generated from a particular choice for LH and RH hypercharges (see section \ref{Sub:Gauge}).  

Once more, we choose to charge RH fields in order to preserve SM-like (for $s_\theta = 0$) the Z boson  interactions with LH fields. From the phenomenological point of view,  we also opted for charging the second generation only. Apart from the anomalous magnetic moment, both for electrons and muons, we can include the effect in the proton charge radius measured from the Lamb shift in e-hydrogen and $\mu$-hydrogen among the most stringent constraints to the model \cite{TuckerSmith:2010ra}. 

In this framework, LFV will be mediated by the currents
\begin{equation}
\mathcal{L} \supset c_\theta  g_X  \ X^\mu \ \overline{l}_{R i} \gamma_\mu [X \delta^{i2} \delta^{j2}] l_{R j} .
\end{equation}
When $e_R$ represents the mass eigenstates, such that $e_R = V^l_R l_R$, the term is converted to 
\begin{equation}
\mathcal{L} \supset c_\theta  g_X  \ X^\mu \ \overline{e}_{R a} \gamma_\mu (V^l_R )_{ai} [X \delta^{i2} \delta^{j2}] (V^l_R)^\dagger_{jk} e_{R k} 
\rightarrow 
c_\theta  g_X  X^\mu \ \overline{e}_{R a} \gamma_\mu  [X (V^l_R )_{a2} (V^l_R)^*_{k2}]  e_{R k} .
\end{equation} 
Therefore, the matrices introducing flavor violation and non-universality are given by $\mathbb{F}_{ij} = (V_R )_{i2} (V_R)^*_{j2}$. We see that only a coherent explanation about the possible alignment between flavor and mass eigenstates could confirm the assumption of small FCNC processes in the model. Some hints in different sectors of the Lagrangian could be selected, for instance, through the CKM matrix, defined as
\begin{equation}
(V_{CKM})_{ij} = (V_L^U)_{ik} (V_L^D)^*_{jk}, 
\end{equation}   
with implicit summation on $k$. Thus, if the mass and flavor eigenstates were approximately aligned, simultaneously for U- and D-type LH quarks, all the non-diagonal (ND) elements of $V_{CKM}$ would be suppressed compared to the diagonal ones. The inverse, however, is not true - the presence of phases could suppress ND elements of the CKM but with large terms in the above summation facing a negative interference. Phenomenologically, since the Wolffenstein parameter $\lambda \simeq 0.22$, ND CKM terms are in fact smaller. Once we do not have any particular information on the matrices  components, nothing can be affirmed about interferences or suppression. The same holds, for example, for the PMNS matrix.

In the following we consider $U(1)_X$ models, at low-energies (Mev-GeV), originally proposed to explain  the muon anomalous magnetic moment discrepancy and the proton charge radius discrepancy (see e.g. \cite{Pospelov:2008zw,Barger:2010aj}). These anomalies are often treated as  a signal of lepton flavor non-universality. Thus, in the framework described above, we focus on  the $g_X$ component of the $g_R$ coupling.

\subsection{Gauge interactions with fermions}\label{Sub:Gauge}
The gauge interactions with fermions are described by:
\begin{equation}\label{Eq.GI}
\mathcal{L}_{kin} \supset i \biggl[ \overline{L}_{\alpha L} \slashed{D} {L}_{\alpha L} 
+
\overline{Q}_{\alpha L} \slashed{D} {Q}_{\alpha L} 
+ 
\overline{l}_{\alpha R} \slashed{D} {l}_{\alpha R} 
+ 
\overline{d}_{\alpha R} \slashed{D} {d}_{\alpha R} 
+ 
\overline{u}_{\alpha R} \slashed{D} {u}_{\alpha R} 
+ 
\overline{\chi}_{ R} \slashed{D} {\chi}_{ R} \biggr], 
\end{equation}
with $\alpha = 1,2,3$. In terms of mass eigenstates the covariant derivative can be written as
\begin{equation}
D_\mu = \partial_\mu - i g(W^+ \mathbb{I}_+ + W^- \mathbb{I}_-) -i e \mathbb{Q} A_\mu -i  g_Z Z_\mu
-i g_R X_\mu
\end{equation}
and
\begin{eqnarray}
e \mathbb{Q} &=&  g s_\phi \tau^3 + g_Y c_\phi Y ,\nonumber \\
g_Z &=& c_\theta g_Z^{SM} + s_\theta (\kappa Y + g_X X) 
= g_Z^I + s_\theta g_X X ,\nonumber \\
g_R &=& s_\theta g_Z^{SM} - c_\theta (\kappa Y + g_X X)
= g_R^I - c_\theta g_X X\, ,
\end{eqnarray}
where the coupling constant components proportional to the identity - i.e. those depending only on the charges assignment under the SM gauge group - have been separated and describe flavor universal vertexes: 
\begin{equation}
g_Z^I \equiv c_\theta g_Z^{SM} + s_\theta \kappa Y, 
\qquad
g_R^I \equiv s_\theta g_Z^{SM} - c_\theta \kappa Y\,.
\end{equation}

The charged currents occur entirely like in the SM, weighted by the CKM matrix. 
The new physics (NP) effects are limited to the neutral currents. The parameter $s_\theta$ introduces the size of NP contributions in comparison with   the SM ones. Therefore, it can be  a small parameter. Since we choose to charge RH fields only, the amount of flavor violating processes in both $Z$ and $X$ interactions is related to the $g_X X_\mu$ term and is therefore exclusive to RH sector, taking place in the second generation. Defining the vector of fermion fields $f = (f_1, f_2, f_3)$ and rotating the system to the mass basis, $f_R \rightarrow V_{fR} f_R' \equiv  V_{fR} f_R$, the general currents depending on the $X$ charges can be fully separated via:
\begin{eqnarray}\label{Eq.GIb}
\mathcal{L}_{kin} &\supset&  \biggl[ g_R^I(u_R) \ \overline{u}_{R} \gamma^\mu {u}_{R} 
+  g_R^I(d_R) \ \overline{d}_{R}  \gamma^\mu {d}_{R} 
+  g_R^I(l_R) \ \overline{l}_{R} \gamma^\mu {l}_{R}
\biggr] X_\mu \nonumber \\
& & - c_\theta g_X \biggl[ \overline{u}_{R} \mathbb{F}^U \gamma^\mu {u}_{R} 
\ + \overline{d}_{R} \mathbb{F}^D \gamma^\mu {d}_{R} 
\ + \overline{l}_{R} \mathbb{F}^l \gamma^\mu {l}_{R}
\biggr] X_\mu .
\end{eqnarray}
The matrices \begin{equation}
\mathbb{F}^f \equiv V^\dagger_{fR} \mathbb{X}^f V_{fR}, \qquad \text{where} \qquad 
(\mathbb{X}^f)_{ij} \equiv X^f \delta_{2i} \delta_{2j} ,   
\end{equation} or
\begin{equation}
(\mathbb{F}^f)_{ij} = X^f (V^\dagger_{fR})_{i2} (V_{fR})_{2j}, 
\end{equation}
summarize the amount of flavor violation and fermion non-universality in the model. Once more, in the scenario where flavor is aligned to the mass eigenstates, i.e. when the absolute value of diagonal elements of $V_{f R}$ are larger than the non-diagonal ones, the flavor violating processes must favor the second generation in the final state, since it would include at least one factor of $(V_{f R})_{22}$. Moreover, the diagonal elements also fix the amount of LFV by
\begin{equation}
\mathbb{F}^f = X^f \begin{pmatrix}
|V_{f R}|^2_{21} & (V^\dagger_{fR})_{12} (V_{fR})_{22} & (V^\dagger_{fR})_{12} (V_{fR})_{23} \\ (V^\dagger_{fR})_{22} (V_{fR})_{21} & |V_{f R}|^2_{22} & (V^\dagger_{fR})_{22} (V_{fR})_{23} \\ (V^\dagger_{fR})_{32} (V_{fR})_{21} & (V^\dagger_{fR})_{32} (V_{fR})_{22} & |V_{f R}|^2_{23}
\end{pmatrix}, 
\end{equation} 
or, one can define,
\begin{equation}
|\mathbb{F}^f| \equiv X^f \begin{pmatrix}
|V_{f R}|^2_{21} & |V_{fR}|_{21} |V_{fR}|_{22} & |V_{fR}|_{21} |V_{fR}|_{23} \\ |V_{fR}|_{21} |V_{fR}|_{22} & |V_{f R}|^2_{22} & |V_{fR}|_{22} |V_{fR}|_{23} \\ |V_{fR}|_{21} |V_{fR}|_{23} & |V_{fR}|_{22} |V_{fR}|_{23} & |V_{f R}|^2_{23}
\end{pmatrix}\,. 
\end{equation}
Due to  unitarity of $V_{fR}$, the trace of $\mathbb{F}^f$ is equal to $X^f$:
\begin{eqnarray}
\text{Tr}[\mathbb{F}^f] &=& \text{Tr}[ V^\dagger_{fR} \mathbb{X}^f V_{fR}] 
= \text{Tr}[\mathbb{X}^f] 
 =X^f\,. 
\end{eqnarray}
Naturally, the closer one of the diagonal entries is to $X_f$, smaller the  LFV is  predicted by the model. Notice that, unlike the CKM matrix, $\mathbb{F}$ does not enclose all the physical processes involving RH fields and the matrices $V_R$ must be independently present in the scalar interactions.

The interaction with $Z$ follows a similar pattern but it is  doubled suppressed by $s_\theta g_X \approx g_X^2$, i.e. flavor changing and non-universality are dominated by $X^\mu$ interactions. All the non-diagonal vertexes are summarized in the second line of Eq.(\ref{Eq.GIb}), represented by the matrix $\mathbb{F}$, and it 
is  useful to separate the diagonal currents in a simplified form. Here these terms will be written like 
\begin{equation}\label{Eq.NeutralCur}
\mathcal{L} \supset \frac{1}{2} \ \overline{f} \ \gamma_\mu (g_V^f + g_A^f \gamma^5) \ f \ Z^\mu
+ \frac{1}{2} \ \overline{f} \ \gamma_\mu (x_V^f + x_A^f \gamma^5) \ f \ X^\mu, 
\end{equation}
such that
\begin{eqnarray}
x_V^f &=& g_R^I (f_R) + g_R^I (f_L) - c_\theta g_X \mathbb{F}^f_{ii},   \\
x_A^f &=& g_R^I (f_R) - g_R^I (f_L) - c_\theta g_X \mathbb{F}^f_{ii}.
\end{eqnarray}
By replacing the electric charges and hypercharges:
\begin{subequations}\label{Eq.coup}
	\begin{eqnarray}
x_V^U &=&  g \frac{s_\theta}{c_\phi} \left(\frac{1}{2} - \frac{4}{3} s^2_\phi \right)
- c_\theta \kappa \frac{5}{6} - c_\theta g_X \mathbb{F}^U_{ii},
\\
x_A^U &=&  g \frac{s_\theta}{c_\phi} \left(- \frac{1}{2} \right)
- c_\theta \kappa \frac{1}{2} - c_\theta g_X \mathbb{F}^U_{ii},
\\
x_V^D &=&  g \frac{s_\theta}{c_\phi} \left(- \frac{1}{2} + \frac{2}{3} s^2_\phi \right)
+ c_\theta \kappa \frac{1}{6} - c_\theta g_X \mathbb{F}^D_{ii},
\\
x_A^D &=&  g \frac{s_\theta}{c_\phi} \left(\frac{1}{2}\right)
+ c_\theta \kappa \frac{1}{2} - c_\theta g_X \mathbb{F}^D_{ii},
\\
x_V^l &=&  g \frac{s_\theta}{c_\phi} \left(- \frac{1}{2} + 2 s^2_\phi \right)
+ c_\theta \kappa \frac{3}{2} - c_\theta g_X \mathbb{F}^l_{ii} \label{Eq.xvl}
\\
x_A^l &=& g \frac{s_\theta}{c_\phi} \left(\frac{1}{2}\right)
+ c_\theta \kappa \frac{1}{2} - c_\theta g_X \mathbb{F}^l_{ii},
\\
x_V^\nu &=&  - x_A^\nu = g \frac{s_\theta}{c_\phi}  \left(\frac{1}{2}\right)
+ c_\theta \kappa , \label{Eq.xvnu}
\\
x_V^\chi &=& x_A^\chi = c_\theta g_X.
\end{eqnarray}
\end{subequations}


Clearly, the root for $x_A^l = 0$ results in the purely vectorial leptonic vertexes. In addition, by charging LH currents one may generate LH FCNC bi-linears in Eq.(\ref{Eq.GIb}) and enable the effective operators favored by the $R_{K^{(*)}}$ flavor anomalies \cite{Hiller:2017bzc}. 

In our model we emphasize the interactions mediated by a light $X^\mu$ ($m_X \sim 10^2$ MeV). If compared to the dark vector exchange, the effects from the remaining new fields, like $\overline{H}, \overline{H}_s, \chi_r^0, \overline{\phi}^+$ presented in section \ref{Ap.Rxi}, are negligible due to their presence in the decoupling limit \cite{Bhatia:2017tgo}. In the second part of this work, the free parameters coming from Yukawas are constrained by experimental bounds and not fixed along the analysis.

\paragraph{\texorpdfstring{$X_\mu$ interactions with charged hadrons}{}}
In order to calculate the contribution coming from  the inner $X$-bremsstrahlung from a charged hadron, one must first  perform the transformation 
\begin{equation}\label{Eq:Red1}
B^Y_{\mu} \rightarrow B^Y_{\mu} + \epsilon B^X_{\mu}, 
\end{equation}
which converts the QED covariant derivative
\begin{equation}
D_\mu = \partial_\mu -i e q A_\mu ,
\end{equation}
into a minimal coupling including the vector $X_\mu$. Note that the photon field can always be written as  $A = s_\phi W^3 + c_\phi B^Y$, once the rotation in Eq.(\ref{Eq:Red1}) does not modify the Weinberg angle dependence in terms of the weak couplings\footnote{In other words the resultant shift in Eq.(\ref{Eq.Had})  does not depend if the rotation is performed before or after symmetry breaking.}. Thus
\begin{eqnarray}\label{Eq.Had}
D_\mu &\stackrel{(\ref{Eq:Red1})}{\rightarrow}& \partial_\mu -i e q A_\mu - i e q c_\phi \epsilon B_\mu^X \nonumber \\
 &=& \partial_\mu -i e q A_\mu - i e q c_\phi \epsilon (\mathbb{V}^\intercal_{3i} \mathbf{B}_i) \nonumber \\
 &=& \partial_\mu -i e q A_\mu + i q c^2_\phi \kappa X_\mu + \mathcal{O}^2 , 
\end{eqnarray}
where $\kappa = g_Y \epsilon$ and $c_\theta \approx 1$. The remaining terms include a Z interaction suppressed at second order in the small parameters.

\subsection{Scalar potential}\label{Sec.Pot}
The scalar potential  has the same features as the the given  in Ref. \cite{Bhatia:2017tgo}  (see Eq. (2.26)),   
 e.g. the absence of pseudo-Goldstone bosons ensured by the $\mu$-dependent cubic coupling below:
\begin{align}\label{Eq.pot}
V(\phi_0, \phi_X, s) &= \mu_0 (\phi_0^\dagger \phi_0) + \mu_X (\phi_X^\dagger \phi_X)
+ \mu_s (s^* s) + \lambda_0 (\phi_0^\dagger \phi_0)^2 +
\lambda_X (\phi_X^\dagger \phi_X)^2 +
\lambda_s (s^* s)^2 +
\nonumber \\
&  + \lambda_3 (\phi_0^\dagger \phi_0)(\phi_X^\dagger \phi_X) + \lambda_4 (\phi_0^\dagger \phi_X)(\phi_X^\dagger \phi_0)
+ \lambda_{0s} (\phi_0^\dagger \phi_0) (s^* s) +
\lambda_{0X} (\phi_X^\dagger \phi_X) (s^* s) - \nonumber \\
&  - \mu [(\phi_X^\dagger \phi_0)s + h.c.] .
\end{align}
It is convenient to consider  a gauge-fixing Lagrangian before we  analyze the physical spectra, since it can provide useful tools for the diagonalization of the potential.

\subsubsection{\texorpdfstring{$R_\xi$}{} gauges}
The longitudinal components of the $W$ boson arise from the contributions of both Higgs doublets. The charged scalars will mix due to the Lagrangian 
\begin{equation}
\mathcal{L}_S \supset \dfrac{i}{2} g \left[v_0 (\partial \varphi_0^+ W^-  
- \partial \varphi_0^- W^+) 
+ v_X (\partial \varphi_X^+ W^-  
- \partial \varphi_X^- W^+) \right] .
\end{equation}
The gauge-fixing Lagrangian can be chosen as 
\begin{equation}
\mathcal{L}_{g.f.} \supset - \dfrac{1}{\xi_W} 
\left(\partial W^+ -i g \dfrac{\xi_W}{2}(v_0 \varphi_0^+ + v_X \varphi_X^+) \right)
\left(\partial W^- -i g \dfrac{\xi_W}{2}(v_0 \varphi_0^- + v_X \varphi_X^-) \right). 
\end{equation} 
This Lagrangian produces, for instance, the term
\begin{equation}
\mathcal{L}_{g.f.} \supset - \dfrac{i}{2} g \left[v_0 \partial W^+ \varphi_0^-  
+ v_X \partial W^+ \varphi_X^- \right] 
\end{equation} 
such that, after an integration by parts, it must cancel the equivalent piece in $\mathcal{L}_S$. The remaining terms are
\begin{equation}
\mathcal{L}_{g.f.} \supset - \dfrac{1}{\xi_W} (\partial W^+) (\partial W^-)
- \xi_W \frac{g^2}{4} (v_0 \varphi_0^+ + v_X \varphi_X^+)(v_0 \varphi_0^- + v_X \varphi_X^-).
\end{equation} 
The determinant of the mixing matrix above is obviously zero. The zero eigenvalue is linked to a physical charged scalar and the non-zero one to a Goldstone boson with mass $m^2_\phi = \xi_W \frac{g^2 v^2}{4} =\xi_W m^2_W$.  The scalars mass matrix  $(\mathbb{M}^2_{W})_\xi$ is given in the basis ($\varphi_0^+, \varphi_X^+$) and can be  diagonalized by matrix $\mathbb{R}_\xi^W$ 
\begin{equation}
(\mathbb{M}^2_{W})_\xi \propto 
\begin{pmatrix}
v_0^2 & v_0 v_X \\ v_0 v_X & v_X^2 
\end{pmatrix}, 
\qquad
\mathbb{R}_\xi^W = \frac{1}{v} \begin{pmatrix}
v_X & - v_0 \\ v_0 & v_X
\end{pmatrix},
\end{equation}
such that $[\mathbb{R}_\xi^W] (\mathbb{M}^2) [\mathbb{R}_\xi^W]^\intercal = \begin{pmatrix}
0 & 0\\0 & v^2
\end{pmatrix}$ and $\begin{pmatrix}
\bar{\varphi} \\ \bar{\varphi}_g 
\end{pmatrix} = \mathbb{R}_\xi^W \begin{pmatrix}
\varphi_0^+ \\ \varphi_X^+
\end{pmatrix}
$. 

The Goldstone theorem implies that the matrix $\mathbb{R}_\xi^W$  must be orthogonal to the mixing matrix derived from the potential. Hence, they can be simultaneously diagonalized and, as discussed in the following section, the matrices $\mathbb{R}_\xi$ may be sufficient to diagonalize the entire system. 

The construction  of a gauge fixing  Lagrangian for the neutral $Z$ and $X$ bosons is not as straightforward as in the previous example, although it is based on the same procedure. The  difference comes from the  the mixing matrix which contains two dependent parts on the gauge fixing parameters. In other words, there must be a set of bi-linears in the Z and X component of the following Lagrangian
\begin{eqnarray}
\mathcal{L}_S &=& \left(\sum_{k} z_k \ \partial_\mu \chi_k \right) Z^\mu + 
\left(\sum_{k} x_k \ \partial_\mu \chi_k\right) X^\mu , 
\end{eqnarray}
where $z_k \equiv g_Z(\chi_k) v_k$ and $x_k \equiv g_R(\chi_k) v_k$, with the index being  $k \in [\chi_0, \chi_X, s]$. The masses can likewise be written as $m^2_Z = \sum_{k} z^2_k$ and $m^2_X = \sum_{k} x^2_k$.
After introducing the gauge-fixing part, the scalars shall mix analogously to the charged case, but here with both $Z$ and $X$ components  containing  independent $\xi$ parameters. The mass matrices of both gauge bosons 
are not commuting and the total mass matrix must be diagonalized at once, such that the two non-zero eigenvalues embody the gauge-fixing parameters. Explicitly, we see that the matrix is given by
\begin{equation}\label{Eq.gfNB}
\mathcal{L}_{g.f.} \supset - \frac{\xi_Z}{2}\left(\sum_{k} z_k \ \chi_k \right)^2 - 
\frac{\xi_X}{2} \left(\sum_{k} x_k \ \chi_k\right)^2 .
\end{equation}
In the absence of $\xi_X$, for instance, the expression results in just one non-zero eigenvalue as $\mathcal{L}_{g.f.} \supset \frac{\xi_Z}{2} m_Z^2 \chi^2_Z$, and vice-versa. Therefore, we can write this part of the gauge fixing Lagrangian using two scalars $ \chi_p$ and   $ \chi_a$
\begin{equation}
\mathcal{L}_{g.f.} \supset \frac{M^2_p (\xi_Z, \xi_X)}{2}  \chi^2_p 
+ \frac{M^2_a (\xi_Z, \xi_X)}{2}  \chi^2_a ,
\end{equation}
such that 
\begin{equation}
M^2_p (0, \xi_X) = \xi_X  m_X^2,\qquad
M^2_p (\xi_Z, 0) = \xi_Z  m_Z^2,\qquad
M^2_a (\xi_Z, 0) = M^2_a (0, \xi_X) = 0. 
\end{equation}
The remaining scalar is physical, whose mass is determined from the potential and will be denoted as $\chi_r$. The matrices coming from Eq.(\ref{Eq.gfNB}) are orthogonal to the mixing matrix coming from the potential. Finally, the gauge fixing Lagrangian can be chosen as
\begin{equation}
\mathcal{L}_{g.f.} \supset - \frac{1}{2\xi_Z} \left( \partial Z - \xi_Z \sum_{k} z_k \ \chi_k \right)^2 
- \frac{1}{2\xi_X} \left( \partial X - \xi_X \sum_{k} x_k \ \chi_k \right)^2. 
\end{equation}

\subsubsection{Scalar spectra}\label{Ap.Rxi}
As elaborated in the previous section, the theory contains four Goldstones $\phi_g^+, \phi_g^-, \chi_p, \chi_a$, three physical (pseudo) scalars ($\phi^+, \phi^-, \chi_r$) and three Higgses ($H, \overline{H}, \overline{H}_s$), corresponding to the ten original degrees of freedom.
The vacuum stability equations are extracted from the linear terms of the real neutral scalars $(H_\chi, H_0, H_s)$ and lead to the  conditions:
\begin{itemize}
        \item $H_0$:
	\begin{equation}
	\mu_0 + \lambda_0 v_0^2 + \frac{\lambda_{0s}}{2} v_s^2 - \frac{\mu}{\sqrt{2}} \frac{v_s v_X}{v_0} + \frac{v_X^2}{2} (\lambda_3 + \lambda_4) = 0,\label{VCS1}
	\end{equation}
	\item $H_\chi$:
	\begin{equation}
	\mu_\chi + \lambda_\chi v_X^2 + \frac{\lambda_{\chi s}}{2} v_s^2 - \frac{\mu}{\sqrt{2}} \frac{v_s v_0}{v_X} + \frac{v_0^2}{2} (\lambda_3 + \lambda_4) = 0,\label{VCS2}
	\end{equation}
	\item $H_s$:
	\begin{equation}
	\mu_s + \lambda_s v_s^2 + \frac{\lambda_{0 s}}{2} v_0^2 + \frac{\lambda_{0 \chi}}{2} v_X^2 	- \frac{\mu}{\sqrt{2}} \frac{v_0 v_X}{v_s} = 0. \label{VCS3}
	\end{equation}
\end{itemize}

\paragraph{Mass matrix - charged scalars} 

In the basis $(\phi_0^+, \phi_\chi^+)$, using  the vacuum stability equations (\ref{VCS1}),  (\ref{VCS2}) and  (\ref{VCS3}), the squared mass matrix of charged scalars can be written as
\begin{equation}
\mathbb{M}_{W}^2 = \lambda^+ 
\begin{pmatrix}
v_X^2 & - v_0 v_X \\ - v_0 v_X & v_0^2 
\end{pmatrix},
\qquad \text{where} \quad \lambda^+ \equiv \left(\frac{\mu v_s}{\sqrt{2} v_0 v_X} - \frac{\lambda_4}{2}\right)
\end{equation}
and such that 
\begin{itemize}
	\item Orthogonality: $\mathbb{M}_{W}^2 \cdot (\mathbb{M}_{W}^2)_\xi = (\mathbb{M}^2_{W})_\xi \cdot \mathbb{M}_{W}^2 = 0$,
	\item Diagonalization via $\mathbb{R}_\xi^{W}$
	\begin{equation}
	\mathbb{R}_\xi^{W} \mathbb{M}_{W}^2 (\mathbb{R}_\xi^{W})^\intercal = 
	\begin{pmatrix}
	v^2 & 0 \\ 0 & 0 
	\end{pmatrix}.
	\end{equation}
\end{itemize}
Therefore, there is a charged scalar $\overline{\phi}^+$ such that 
\begin{equation}
m^2_{+} = \lambda^+ v^2.
\end{equation}

\paragraph{Neutral scalars}

First, we write the mass matrix generated by the gauge-fixing part of the Lagrangian in the following form
\begin{equation}\label{Eq.0xi}
\mathbb{M}_{\xi \chi}^2 = \xi_Z \left(\sum_{i} z_i \chi^i \right)^2  + \xi_X \left(\sum_{i} x_i \chi^i \right)^2,
\end{equation}
where $z_i = g_Z(\chi^i) v_i$, $x_i = g_X(\chi^i) v_i$ and $i \in (0, X, s)$. On the other hand, from the potential and the vacuum stabiltiy conditions, in the basis $(\chi^0, \chi^X, \chi^s)$, it follows that
\begin{equation}\label{Eq.0chi}
\frac{\mathbb{M}^2_\chi}{2} = \frac{\mu}{2 \sqrt{2}}
\begin{pmatrix}
\frac{v_s v_X}{v_0} & - v_s & -v_X \\
- v_s & \frac{v_0 v_s}{v_X} & v_0 \\
- v_X & v_0 & \frac{v_0 v_X}{v_s}
\end{pmatrix}.
\end{equation} 
A cross-check can be performed by using the orthogonality of the above matrices. The mass of the physical scalar is given by
\begin{equation}
m^2_{\chi r} = \frac{\mu}{\sqrt{2}} \left(\frac{v_0 v_s}{v_X} + \frac{v_0 v_X}{v_s} + \frac{v_X v_s}{v_0}\right). 
\end{equation}  
i.e. in the absence of a term breaking the residual $U(1)$ symmetry  of the potential, a massless pseudo-Goldstone boson would survive at tree-level in the model. 

Once the matrices in Eq.(\ref{Eq.0xi}) and Eq.(\ref{Eq.0chi}) commute, a particular matrix $\mathbb{R}^\xi$, rotating the gauge-fixing Lagrangian, can be further applied to $\mathbb{M}_\chi$ and will result in a block-diagonal matrix, then diagonalized via a second $\mathbb{R}^V$.
In other words, the total mixing matrix can be written like
\begin{equation}
\begin{pmatrix}
\chi^p \\ \chi^a \\ \chi_r 
\end{pmatrix} 
= 
\mathbb{R}^\xi \mathbb{R}^V 
\begin{pmatrix}
\chi_0 \\ \chi_X \\ \chi_s
\end{pmatrix}. 
\end{equation}
This assertion is proved by considering two real and symmetric commuting matrices $\mathbb{A}, \mathbb{B}$. If $\mathbb{D}$ diagonalizes $\mathbb{A}$, i.e. $\mathbb{D} \mathbb{A} \mathbb{D}^\intercal = \mathbb{X}_A$, it follows that
\begin{equation}
\mathbb{D} [\mathbb{A}, \mathbb{B}] \mathbb{D}^\intercal = 0 \rightarrow 
[\overline{\mathbb{B}}, \mathbb{X}_A] = 0,
\end{equation}
where $\overline{\mathbb{B}} = \mathbb{D} \mathbb{B} \mathbb{D}^\intercal$, or
\begin{equation}
(\overline{\mathbb{B}})_{ik} (\mathbb{X}_A)_{kj} - (\mathbb{X}_A)_{ik} (\overline{\mathbb{B}})_{kj} = 0 \ \rightarrow \ 
(\overline{\mathbb{B}})_{ik} \delta_{kj} (\mathbb{X}_A)_{jj} - \delta_{ik} (\mathbb{X}_A)_{ii} (\overline{\mathbb{B}})_{kj}  = 0 ,
\end{equation}
and finally, 
\begin{equation}
(\overline{\mathbb{B}})_{ij} [(\mathbb{X}_A)_{ii} - (\mathbb{X}_A)_{jj}] = 0. 
\end{equation}
Therefore $\overline{\mathbb{B}}$ is block-diagonal whose dimension is given by the eigenvalues degree of degeneracy. 

\paragraph{The Higgs}

For the real scalars, the vacuum stability  equations  lead to the following matrix  
\begin{equation}
\mathbb{M}^2_H = 
\begin{pmatrix}
2 \lambda_0 v_0^2 + \frac{\mu}{\sqrt{2}} \frac{v_s v_X}{v_0} & v_0 v_X (\lambda_3 + \lambda_4) - \frac{\mu v_s}{\sqrt{2}} & v_0 v_s \lambda_{0s} - \frac{\mu v_X}{\sqrt{2}} \\
v_0 v_X (\lambda_3 + \lambda_4) - \frac{\mu v_s}{\sqrt{2}} & 2 \lambda_X v_X^2 + \frac{\mu}{\sqrt{2}} \frac{v_s v_0}{v_X} & v_X v_s \lambda_{Xs} - \frac{\mu v_0}{\sqrt{2}} \\
v_0 v_s \lambda_{0s} - \frac{\mu v_X}{\sqrt{2}} & v_X v_s \lambda_{Xs} - \frac{\mu v_0}{\sqrt{2}} & 2 \lambda_s v_s^2 + \frac{\mu}{\sqrt{2}} \frac{v_0 v_X}{v_s}  
\end{pmatrix},
\end{equation}
which can be further simplified once we assume the new scales $\mu \sim v_s \gg v$. Hence $(\mathbb{M}_H)_{11} \sim (\mathbb{M}_H)_{22}$ and $(\mathbb{M}_H)_{13} \sim (\mathbb{M}_H)_{23}$, where $\lambda_{0s}$ and $\lambda_{Xs}$ are both positive numbers, a condition to leave the potential bounded from below. Therefore, the fields denoted by $\overline{H}, H$ and $H_s$ have their masses given by
\begin{subequations}
	\begin{eqnarray}
	m^2_{\overline{H}} &\approx& 2\lambda_0 v_0^2 - (\lambda_3 + \lambda_4) v_0 v_X + \frac{\mu v_s}{\sqrt{2}} \left(1 + \frac{v_X}{v_0}\right), \\
	m^2_{H} &\approx& 2\lambda_0 v_0^2 + (\lambda_3 + \lambda_4) v_0 v_X + \frac{\mu v_s}{\sqrt{2}} \left(\frac{v_X}{v_0} - 1\right) ,\\
	m^2_{H_s} &\approx& 2\lambda_s v_s^2 + \frac{\mu}{\sqrt{2}} \frac{v_0 v_X}{v_s} .
	\end{eqnarray}
\end{subequations}
Here the choice for the indexes is motivated by the region where $v_0 \sim v_X$, such that
\begin{equation}
m^2_{\overline{H}} \approx  \sqrt{2} \mu v_s ,\quad
m^2_{H} \approx  (2 \lambda_0 + \lambda_3 + \lambda_4) v_0^2 , \quad
m^2_{H_s} \approx 2 \lambda_s v_s^2  . \quad
\end{equation} 

\subsection{Yukawa Lagrangian}
The choice of  non-universal charges requires the inclusion of at least one additional SM-like Higgs, a necessary condition to obtain the correct fermion mass spectra. In addition,  one cannot preserve the quarks neutral under $X$. 
Actually, we avoid adding a third scalar doublet by assigning the same hypercharge to two fermion generations. Following the proposal of Ref.\cite{Babu:2017olk}, we charge one generation whilst the remaining ones remain neutral. This approach reduces the flavor matrix $\mathbb{F}$ to its minimal version. Therefore, here the new Higgs doublet, charged both under $X$ and $Y$, fills the second Yukawa matrix column of a $U(1)_X$ specific to the second fermion family. The Yukawa Lagrangian in the quark sector, from the notation of Ref.\cite{Bhatia:2017tgo}, can be given by
\begin{align}\label{Eq.Yuk}
	\mathcal{L}_Y &= (\bar{Q}_L)_i \mathbb{Y}^U_{ij} \phi_0^c (U_{R})_j +
	(\bar{Q}_L)_i \mathbb{Y}^D_{ij} \phi_0 (D_{R})_j +
	\nonumber \\
	&  + (\bar{Q}_L)_i \mathbb{Y}^U_{i2} \phi_X^c (U_{R})_2 +
	(\bar{Q}_L)_i \mathbb{Y}^D_{i2} \phi_X (D_{R})_2 , 
\end{align}
with $i \in [1,2,3]$ and $j \in [1,3]$. Note that $X_U + X_D = 2 X_Q$ is valid per generation.

In general, $U(1)_X$ models preserve total equivalence in their gauge structure and are distinguished by the functional form of the equations (\ref{Eq.coup}) under the kinetic and gauge couplings. In other words, the gauge sector can generically be represented by Eq.(\ref{Eq.NeutralCur}) such that a particular X-hypercharge assignment will then be converted into the independent functions defining the fermion couplings of (\ref{Eq.coup}). The version we have selected, for instance, generates at tree-level the same set of vertexes as that of \cite{Babu:2017olk}. As in their case, we request a new singlet scalar $s$ living at the high scale and it might generate 
Majorana mass terms for neutrinos. In addition, it explicitly breaks a residual global $U(1)$ in the potential. Although we aim to favor ``muon-specific'' processes, our  X-hypercharge 
choice  $X_{q_i} = X_{l_i} =0$ for $ i = 1,3$, $X_s = - X_c = 1$ and $X_\mu = - X_{\nu_\mu} = 1$, will still disperse its effects into different flavors. Notwithstanding, under the same constraints, the two models are expected to produce completely independent parameter spaces. In the next part of this work we show the comparison between dark photons and $Z'$ physics (see section \ref{Sec:DPandZ}) facing the same and most stringent bounds in the MeV regime, coming from  the electron and muon anomalous magnetic moment \cite{Pospelov:2008zw} and the neutrino trident production \cite{Altmannshofer:2014pba}. 

The mass Lagrangian (\ref{Eq.Yuk}) can be written in terms of flavor vectors as 
\begin{equation}
\mathcal{L}_{mass} \supset \frac{v}{\sqrt{2}} [\overline{u}_L (\mathbb{Y}_0^U c_\beta + \mathbb{Y}_X^U s_\beta) u_R
+ \overline{d}_L (\mathbb{Y}_0^D c_\beta + \mathbb{Y}_X^D s_\beta) d_R] + h.c.\, ,
\end{equation}
where $s_\beta = \frac{v_X}{v}$. Again, $\mathbb{Y}_0$ has filled the  first and third columns, while $\mathbb{Y}_X$  has non-zero elements in the second one. 
By rotating the quark vectors as $u_R \rightarrow V_{uR} u'_R \equiv V_{uR} u_R, \ u_L \rightarrow V_{uL} u_L, \ d_R \rightarrow V_{dR} d_R, \ d_L \rightarrow V_{dL} d_L $, it follows
\begin{equation}
	\mathcal{L}_{mass} \supset \frac{v}{\sqrt{2}} [\overline{u}_L \ V^\dagger_{uL}(\mathbb{Y}_0^U c_\beta + \mathbb{Y}_X^U s_\beta)V_{uR} \ u_R
	+ \overline{d}_L \ V^\dagger_{dL}(\mathbb{Y}_0^D c_\beta + \mathbb{Y}_X^D s_\beta)V_{dR} \ d_R] + h.c. \, ,
\end{equation}
with real, non-negative and diagonal matrices defining the quark masses.
For completeness, the interactions with neutral scalars can be put in the form
\begin{equation}
	\mathcal{L}^h \supset \overline{u}_L (\kappa_0^U H^0 + \kappa_X^U H^X) u_R + h.c.\, ,
\end{equation}
where $\kappa_0^U \equiv V^\dagger_{uL} \mathbb{Y}_0^U V_{uR}$ and $\kappa_X^U \equiv V^\dagger_{uL} \mathbb{Y}_X^U V_{uR}$. Since there are no interactions among quarks and the singlet $s$, the vertices can be written as
\begin{equation}
	\mathcal{L}^h \supset \langle \mathbf{j}^U | \mathbf{h} \rangle + h.c.\, ,
\end{equation}
with $\mathbf{j}^U \equiv (j_0^U, j_X^U, 0)$, $\mathbf{h} \equiv (H_0, H_X, H_s)$ and $j_i^U \equiv \overline{u}_L \ \kappa_i^U  u_R$. Thus, in the mass basis,
\begin{equation}
	\mathcal{L}^h \supset \langle \mathbb{R}^h \mathbf{j}^U | \overline{\mathbf{h}} \rangle + h.c.\, .
\end{equation}
The matrix $\mathbb{R}^h$ rotates $\mathbb{M}_h^2$ according to Section \ref{Sec.Pot}. Finally, 
\begin{equation}
	\mathcal{L}^h \supset [(\mathbb{R}^h)_{11} \ j_0^U + (\mathbb{R}^h)_{12} \ j_X^U] \overline{H} 
	+ [(\mathbb{R}^h)_{21} \ j_0^U + (\mathbb{R}^h)_{22} \ j_X^U] H + 
	[(\mathbb{R}^h)_{31} \ j_0^U + (\mathbb{R}^h)_{32} \ j_X^U] \overline{H}_s
	+ h.c.\, .
\end{equation}
The same follows for down-quarks and charged leptons. For instance,
\begin{equation}
	\mathcal{L}_{mass} \supset \frac{v}{\sqrt{2}} [\overline{l}_L (\mathbb{Y}_0^l c_\beta + \mathbb{Y}_X^l s_\beta) l_R] + h.c.\, ,
\end{equation}
The lepton vectors are rotated as $l_R \rightarrow V_{lR}l_R, \ l_L \rightarrow V_{lL}l_L$, such that $V^\dagger_{lL}(\mathbb{Y}_0^l c_\beta + \mathbb{Y}_X^l s_\beta)V_{lR}$ defines their mass matrix.

\subsection{Parameter space}\label{Sec:ParSpc}
The multi-dimensional free parameter space in models beyond the Standard Model is commonly larger than the simple $g_X \times m_X$ planes. In order to avoid a redundant criterion for fixing these planes, it is important to 
consider all relations  emerging in the gauge sector and connecting the remaining variables at tree level.
The procedure is equivalent to the reduction of  the dimension of a multi-variable set through some associated set of independent equations. In fact, a natural relation encompasses coupling constants and energy scales which, in general, may be directly fitted by the observable connected to it. 
In addition to that, one can also permute some of the variables. Such a replacement does not reduce the dimension of parameter space, but it might lead to a more convenient  use of the model. Let us consider the SM example, which is initially described by the $P$ set
\begin{equation}
P := [g, g_Y, v]. 
\end{equation} 
After the $W_3$ - $B$ mixing, the angle parameterizing the eigenvectors can be 
used in place of $g_Y$, i.e.
\begin{equation}
P := [g, g_Y, v] \rightarrow [g, s_w, v]. 
\end{equation}
In all vertexes, $g_Y$ must be written as $g_Y(g, s_w)$ (in fact $g_Y$ does not depend on $v$). Now, from the $Z$ pole mass one can  perform a fit of the parameters which  eliminates, for instance, any dependence on $v$ (i.e. $v = v(g, s_w)$). Thus,
\begin{equation}
P \stackrel{m_Z}{\rightarrow} [g, s_w]. 
\end{equation}
Next, once the charged currents are coupled only through the $g$ coupling, it can be related to  the Fermi constant $G_F$ at  the low energy limit, i.e. $P \rightarrow [g, s_w] \stackrel{G_F}{\rightarrow} [s_w]$. Finally, from the requirement that the theory must reproduce the electromagnetic interactions one last independent equation is given by 
\begin{equation}
g s_w = e, \qquad g_Y(g, s_w) c_w = e\, .
\end{equation}
Therefore, the gauge sector of the Standard Model is  fully determined. 
We must note that the $W$ pole mass was not used in any of the steps  presented above and it emerges as a prediction of the model. 

\vspace{.5cm}
As mentioned before, the $P$ set is defined by the variables entering in the New Physics effective couplings and includes the mixing matrices. In our model, $P$ is necessarily larger than in the SM but it still allows a significant reduction. Initially, it follows that
\begin{equation}
P := [\kappa, g, g_Y, g_X, v_X, v_0, v_s, \mathbb{F}]
\end{equation}
Similarly to the previous example, the constants $g$, $g_Y$ are solved in terms of the remaining elements.  Since in the asymptotic limit  $m_Z$  depends only on $v$ it might be convenient to preserve $c_\beta$ in the analysis. Finally, the $v_s$ breaking scale can be \textit{replaced} by $m_X$. We end up with a five-dimensional parameter space, namely
\begin{equation}
P := [c_\beta, \kappa, g_X, m_X, \mathbb{F}]\,.
\end{equation}
The kinetic mixing constant is independent and may be replaced by the new mixing angle $\theta$. Accordingly, there must be a region for $\kappa$ where the $Z$ interactions are exactly described as in the SM, i.e. where $s_\theta = 0$.

\section{Dark photons vs. \texorpdfstring{$Z'$}{} gauge bosons}\label{Sec:DPandZ}

The full set of dark gauge bosons   $X_\mu$ can be divided into two subsets, namely the one composed out of  dark photons, here denoted by $A'$, coupled exclusively to vector currents. The second subset comprises general $Z'$ bosons whose couplings include axial-vector  components. In the following paragraphs  we briefly summarize the current theoretical status as well as the results of  experimental  searches for the effects of these fields \cite{Feng:2016ysn}.
 
\paragraph{Dark Boson Searches and Future Experiments}
From our study in the Section \ref{Sub:Gauge}, a  general property of the vector and axial-vector couplings  is that  both contain universal and non-universal parts. The LEP searches \cite{Fox:2011fx} can primarily test possible electron couplings to dark fields by looking for recoil energy in a nucleus and therefore can be used to place bounds on the universal part.  On the other hand, experiments such as $Mu3ee$ \cite{Echenard:2014lma}, devoted to test LFV via the decay channel $\mu \rightarrow e e^+ e^-$, can place bounds on the flavor matrix\footnote{Matrix which summarizes the amount of non-universality and flavor violation in the model.} of the particular model and will cover the range $10$ MeV$ < m_{A'}< 80$ MeV. The BaBar collaboration has also performed $A'$ searches \cite{TheBABAR:2016rlg} and their results highly constrain dark photons with mass above the di-muon threshold.  

In the minimal dark photon case, i.e. where the fields are neutral under $X$ and the couplings defined exclusively by the kinetic mixing constant, one can mention the results of the KLOE experiment \cite{Archilli:2011zc}, in which the searches were performed in $\phi \rightarrow \eta X$, $\eta \rightarrow \pi^+ \pi^- \pi^0$, $X \rightarrow e^+ e^-$, with the  dark photon  mass in the range  $50$ MeV $ < m_{A'} < 420$ MeV. The NA$48/2$ collaboration \cite{Batley:2015lha} has covered a $9$ MeV  $< m_{A'}< 120$ MeV  range in kaon decays $K \rightarrow \pi \pi^0$ and $K \rightarrow \mu \nu \pi^0$;

The Run 3 of LHCb plans to search for dark photons in the  charm meson decays $D^* \rightarrow D^0 A' (A' \rightarrow e^+ e^-)$ and is scheduled for 2021-2023 \cite{Ilten:2015hya}. The DarkLight experiment \cite{Balewski:2014pxa} will be sensitive to $10$ MeV$ < m_{A'}  < 100$ MeV by e-H scattering producing on-shell dark photons. It is scheduled for 2018-2020.
Similarly, the Heavy Photon Search \cite{Moreno:2013mja} will scatter an electron beam on a Tungsten target, and is scheduled for 2020. 
Finally, the NA62 experiment \cite{Dobrich:2017yoq} can place bounds on the light $Z'$ couplings by measuring the rate of the rare decays $K \rightarrow \pi \bar{\nu} \nu$.

\begin{figure}[tbp]
	\centering
	\subfigure[]{
		\includegraphics[width=.45\textwidth]{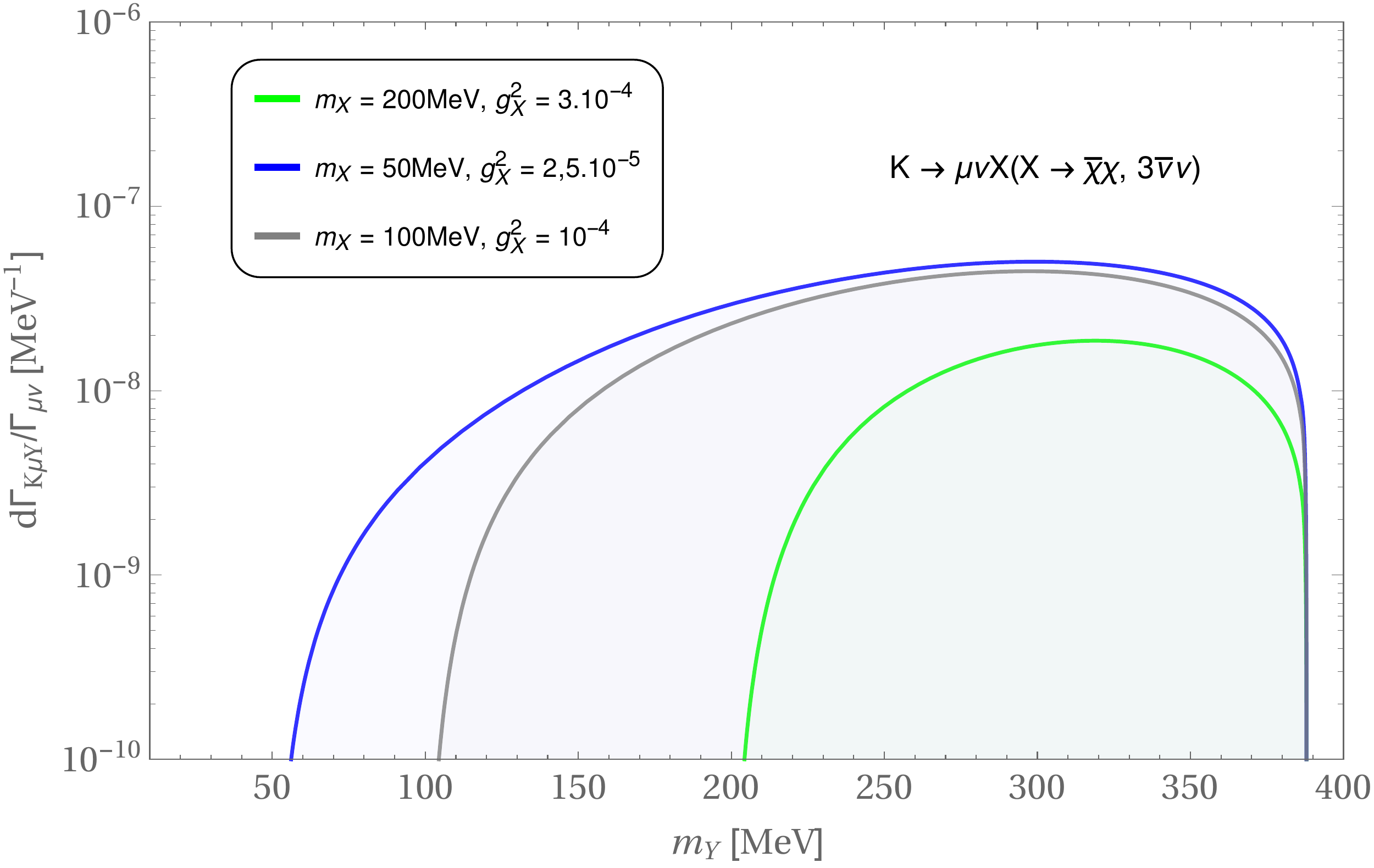}
	}
	\hspace{0.3cm}
	\subfigure[]{
		\includegraphics[width=.45\textwidth]{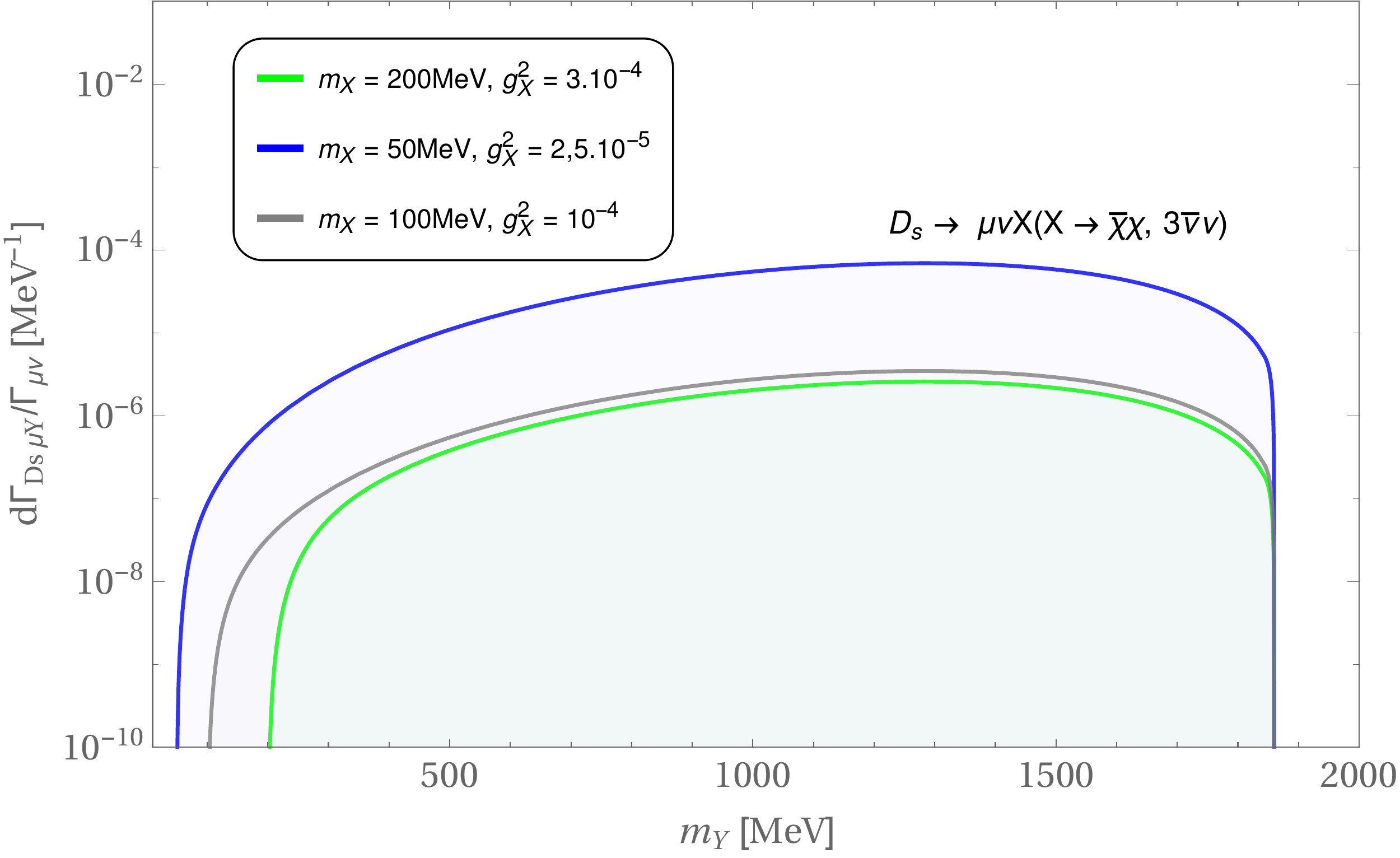}
	}\label{fig:Diff}
	\caption{The Differential decay width $d\Gamma_{M\mu Y}$ normalized  by $\Gamma_{\mu \nu}$ for $M = K, D_s$. The curves are plotted under the fixed values $(c_\beta, \kappa, F_{\mu \mu}, m_\chi) = (0.8, -4g_X, 1, 3m_X)$. In Fig.(b) the channel $D_s \rightarrow \tau \bar{\nu}_\tau (\tau \rightarrow \mu \nu_\tau \bar{\nu}_\mu)$ hides the distribution generated by $X \rightarrow \bar{\chi} \chi, 3\bar{\nu}\nu$.}
\end{figure}

\paragraph{Constraints}
The long-standing discrepancy between the measured and theoretically predicted muon anomalous magnetic moment  \cite{Bennett:2006fi,Bennett:2004pv,Jegerlehner:2009ry,Hagiwara:2011af,Davier:2010nc}  is at the level of  $\sim 3.5-4\, \sigma$. Many approaches of physics beyond the SM were used to resolve this discrepancy by assuming only one new mediating particle \cite{Freitas:2014pua,Queiroz:2014zfa,Biggio:2014ela,Biggio:2016wyy}.
In the  second part of this work we will present  the application of different versions of the present model under the most stringent bounds in the MeV regime. Among these processes we can mention  the electron anomalous magnetic moment \cite{Pospelov:2008zw}, $\nu e$ scattering \cite{Williams:2011qb}, parity non-conserving observables in $Z'$ phenomenology, neutrino trident production \cite{Altmannshofer:2014pba} and the missing energy searches in $K \rightarrow \mu Y$ \cite{Pang:1989ut}. In Fig. \ref{fig:Diff} the differential decay width $d\Gamma_{M\mu Y} / \Gamma_{\mu \nu}$ for $M = K, D_s$ is presented, motivated by the work \cite{Pang:1989ut}, with the fixed values $(c_\beta, \kappa, F_{\mu \mu}, m_\chi) = (0.8, -4g_X, 1, 3m_X)$. In (b) the differential decay width  for  $D_s \rightarrow \tau \bar{\nu}_\tau (\tau \rightarrow \mu \nu_\tau \bar{\nu}_\mu)$ must overshadow the  $d\Gamma_{D_s \mu Y}$ normalized  by $\Gamma_{D_s\to\mu \nu}$. 

In order to maximize the parameter space covered in our   analysis, our strategy includes DM considerations applied to a stable $\chi$ fermion, in principle lighter than $Z'$. In addition, we notice that lepton non-universality in the first and second families will necessarily imply a discrepancy in the proton charge radius estimated from the Lamb shift in the e-hydrogen and $\mu$-hydrogen system \cite{Barger:2010aj}, such that a precise measurement of such processes must correspond to one of the most severe bounds for non-universal dark boson theories.  

Finally, effects of dark fields in purely leptonic processes support DM searches at future lepton colliders.  The same physics may still  give some effects in the leptonic meson decays such as $M \rightarrow \mu \nu ee$ for $M = K, D, D_s, B$. In the subsequent part of this work, we compute the SM  branching ratios for these channels and compare them  to  the results  from  the $Z'$-boson exchange.   
 
\section{Conclusions}
We have seen that the minimal anomaly free $SM \otimes U(1)_X$ models may provide solutions 
to explain existing anomalies in the leptonic sector at low energies.  In the UV complete version
presented here, the quantum anomalies are canceled per generation. The chiral X-hypercharges for a single family require a second Higgs doublet and a scalar singlet in order to provide to the model a consistent fermion mass spectra. Right-handed fermions  are incorporated, being constrained by the neutrino interactions, while by charging the second generation under $ U(1)_X$,  we find  a convenient framework to explain the  discrepancies involving muons.  
Finally, we  considered  $X_\mu$  bosons either as  dark photons $A'$, or $Z'$  gauge bosons, according to the role played by the vector-axial currents, aiming to test the common assertion that $Z'$ physics might be disfavored by parity non-conserving effects. This work is a theoretical  introduction  which will be accompanied by a  more complete phenomenological analysis, developed under the dark matter considerations applied to a light ($MeV$) and stable dark $\chi_R$ fermion. It establishes both the notation and the relations between the new parameters which might be useful for future studies.   

\newpage
\appendix
\section{Appendix}

\subsection{Possibilities for the new fermion \texorpdfstring{$\chi_R$}{} and neutrino masses}\label{Ap.FermionChi}
In the version with Majorana fermions the model must include a new sterile generation such that the mass Lagrangian can be written as:

	\begin{subequations}\label{Eq.Maj1}
		\begin{eqnarray}
		\mathcal{L} &=& - \frac{M^\chi}{2} \left[\overline{(\chi_{R1})^c} \chi_{R 1} + \overline{\chi_{R1}} (\chi_{R1})^c \right]
		\\ 
		& &  - \frac{Y_{1 2}}{\sqrt{2}} \left[\overline{(\chi_{R1})^c} \chi_{R 2} + \overline{(\chi_{R2})^c} \chi_{R 1} \right] s - \frac{Y_{1 2}^*}{\sqrt{2}} \left[\overline{\chi_{R2}} (\chi_{R 1})^c + \overline{\chi_{R1}} (\chi_{R 2})^c \right] s^* \label{Eq.MajYuk}.
		\end{eqnarray}
	\end{subequations}

The brackets in the Yukawa Eq.(\ref{Eq.MajYuk})  lead to   a  symmetric Majorana mass matrix.  The conjugated field is defined by
\begin{equation}
\chi_R^c = C \overline{\chi_R}^\intercal \,, 
\end{equation} 
in terms of a general realization of the charge conjugation operator $C$, such that
\begin{eqnarray}
\chi &\rightarrow& \chi^c = C \overline{\chi}^\intercal = -\gamma^0 C \chi^* \\
\overline{\chi} &\rightarrow& \overline{\chi^c} = -\chi^\intercal C^\dagger\, .
\end{eqnarray}
The Lagrangian can be rewritten like
\begin{eqnarray}
\mathcal{L} &\supset& - \frac{m^\chi}{2} \left[{(\chi_{R1})^\intercal} C^\dagger \chi_{R 1} + {\chi_{R1}^\dagger} C \chi_{R1}^* \right] - 
\\ 
& &  - \frac{Y_{1 2}}{\sqrt{2}} \left[{(\chi_{R1})^\intercal} C^\dagger \chi_{R 2} + {(\chi_{R2})^\intercal} C^\dagger \chi_{R 1} \right] s - \frac{Y_{1 2}^*}{\sqrt{2}} \left[{\chi^\dagger_{R2}} C \chi_{R 1}^* + {\chi^\dagger_{R1}} C \chi^*_{R 2} \right] s^*\,.
\end{eqnarray}
In principle the mass matrix elements are complex numbers. However, in the minimal $2 \times 2$ version a redefinition of the $\chi_{R\alpha}$ fields can absorb their phases, leaving the final matrix real and symmetric:
\begin{equation}
\mathcal{L} \supset \frac{1}{2} \chi^\intercal_{R \alpha} C^\dagger M_{\alpha \beta} \chi_{R\beta} + h.c.\,.
\end{equation}
The mass matrix arises after EWSB, given by
\begin{equation}
\mathbb{M} = \begin{pmatrix}
m & Y_{12} v_s \\ Y_{12} v_s & 0\, ,
\end{pmatrix},
\end{equation}
diagonalized via
\begin{equation}
(V^\chi_R)^\intercal \mathbb{M} (V^\chi_R) = \mathbb{M}^\chi\, ,
\end{equation}
with $V^\chi_R$ being unitary, a criterion to give $\mathbb{M}^\chi$ with real and positive eigenvalues. The fields are rotated  as $\chi_R \rightarrow V_R^\chi \chi'_R \equiv V_R^\chi \chi_R$, where in the r.h.s. the same notation is used for the mass states.
 In conclusion, 
\begin{equation}
\mathcal{L}^\chi \supset \frac{m_k}{2} \chi^\intercal_{R k} C^\dagger \chi_{R k} + h.c.\,.
\end{equation}
$k = 1,2$, or, equivalently,
\begin{equation}
\mathcal{L}^\chi \supset - \frac{m_k}{2} \overline{\chi^c_{R k}} \chi_{R k} + h.c.\,. 
\end{equation}
where the Majorana dark fermions 
\begin{equation}
\chi_k = \chi_{Rk} + \chi^c_{Rk} 
\end{equation}
have well defined mass. Note that these fields are protected to decay due to a global $U(1)^\chi$. The portal into SM is created by both $Z$ and $X$ interactions.

\paragraph{Neutrino masses}
\begin{subequations}\label{Eq.EleLag}
	\begin{eqnarray}
	\mathcal{L} &=& - \frac{1}{4} \mathbf{W}^{\mu \nu} \cdot
	\mathbf{W}_{\mu \nu} 
	- \frac{1}{4} B^{Y \mu \nu} B^Y_{\mu \nu}
	- \frac{1}{4} B^{X \mu \nu} B^X_{\mu \nu} 
	+ \frac{\epsilon}{2} 
	B^{Y \mu \nu} B^X_{\mu \nu} +
	\\ 
	& &  + 	(D_\mu \phi^0)^\dagger (D^\mu \phi^0)
	+
	(D_\mu \phi^X)^\dagger (D^\mu \phi^X) 
	+ 
	(D_\mu s)^\dagger (D^\mu s)
	-
	V(\phi^0,\phi^X,s)
	\\ 
	& & - \sum_{\alpha = 1,2,3} \sum_{\beta = 1,3} \left(\overline{L}_{\alpha L} \phi^0 Y^l_{\alpha \beta} e_{\beta R} + \overline{L}_{\alpha L} \tilde{\phi}^0 Y^\nu_{\alpha \beta} \nu_{\beta R} + h.c.\right)
	\\ 
	& & - \sum_{\alpha = 1,2,3} \left(\overline{L}_{\alpha L} \phi^X Y^l_{\alpha 2} e_{2 R} + \overline{L}_{\alpha L} \tilde{\phi}^X Y^\nu_{\alpha 2} \nu_{2 R} + h.c.\right)
	\\ 
	& & - \sum_{\alpha = 1,2,3} \sum_{\beta = 1,3} \left(\overline{Q}_{\alpha L} \phi^0 Y^D_{\alpha \beta} d_{\beta R} + \overline{Q}_{\alpha L} \tilde{\phi}^0 Y^U_{\alpha \beta} u_{\beta R} + h.c.\right)
	\\ 
	& & - \sum_{\alpha = 1,2,3} \left(\overline{Q}_{\alpha L} \phi^X Y^D_{\alpha 2} d_{2 R} + \overline{Q}_{\alpha L} \tilde{\phi}^X Y^U_{\alpha 2} u_{2 R} + h.c.\right)
	\\ 
	& &  - \sum_{\alpha = 1,3} Y_{\alpha 2} \left[\overline{(\nu_{R\alpha})^c} \nu_{R 2} + \overline{(\nu_{R2})^c} \nu_{R \alpha} \right] s - Y_{\alpha 2}^* \left[\overline{\nu_{R2}} (\nu_{R \alpha})^c + \overline{\nu_{R\alpha}} (\nu_{R 2})^c \right] s^* 
	\\ 
	& & - \sum_{\alpha, \beta = 1,3} \frac{M_{\alpha \beta}}{2} \overline{(\nu_{R\alpha})^c} \nu_{R \beta} + h.c.
	\\ 
	& & + i \sum_{\alpha = 1,2,3} \biggl[ \overline{L}_{\alpha L} \slashed{D} {L}_{\alpha L} 
	+
	\overline{Q}_{\alpha L} \slashed{D} {Q}_{\alpha L} 
	+ \\ 
	& & + \overline{l}_{\alpha R} \slashed{D} {l}_{\alpha R} 
	+ 
	\overline{d}_{\alpha R} \slashed{D} {d}_{\alpha R} 
	+ 
	\overline{u}_{\alpha R} \slashed{D} {u}_{\alpha R} 
	+ 
	\overline{\nu}_{\alpha R} \slashed{D} {\nu}_{\alpha R} 
	\biggr]\, ,
	\end{eqnarray}
\end{subequations}
above, the index ``c'' denotes charge conjugated and is used to recover that Majorana mass terms may appear as long as the charges satisfy $X_\alpha + X_\beta = 0$. In the above variant this piece is composed by the sterile neutrinos of first and third generations. Since $\nu_{2 R}$ is charged under $X$, the mixing with the additional fermions is inserted via the singlet $s$ and under the condition $X_\alpha + X_2 + X_s = 0$.
In this section we  briefly present some general aspects of the see-saw mechanism which concludes the construction of the model. 

By including three generations of RH neutrinos, after the EWSB the Dirac and Majorana mass terms can be included in the Lagrangian  (see \cite{Branco:1999fs})
\begin{equation}
\mathcal{L}_\nu = - \frac{1}{2}
\begin{pmatrix}
\overline{\nu_L^c} &  \overline{\nu_R} 
\end{pmatrix}
\begin{pmatrix}
\mathbb{M}_L & \mathbb{M}_D \\ \mathbb{M}_D & \mathbb{M}_R
\end{pmatrix}
\begin{pmatrix}
\nu_L \\ \nu^c_R
\end{pmatrix} + h.c.\, , 
\end{equation}
which can be in this form due to $\overline{\psi^c} \psi^c = - \psi^\intercal C^{-1} C \overline{\psi}^\intercal = \overline{\psi} \psi$, where in the last step the anti-commuting property of fermion fields has been considered\footnote{This minus sign does not appear in mass terms from the property of hermitian conjugate of Grasmann fields.} (see 3.52 \cite{Branco:1999fs}). Here the matrix $\mathbb{M}_L = 0$ once there is no Majorana mass terms for LH neutrinos. The mass matrix can be rotated by the following transformations
\begin{equation}
U^\dagger M U^* = \begin{pmatrix}
\mathbb{D}_L & \\ & \mathbb{D}_R
\end{pmatrix} \qquad \text{where} \qquad 
U = \begin{pmatrix}
\mathbb{V}_L & \mathbb{V}_{LR} \\ \mathbb{V}_{RL} & \mathbb{V}_R
\end{pmatrix}
\end{equation}
In the framework where $\mathbb{M}_D \ll \mathbb{M}_R$, the matrices $\mathbb{V}_L$ and $\mathbb{V}_R$ may be taken approximately unitary, with $\mathbb{V}_{RL}, \mathbb{V}_{LR}$ suppressed by $\approx \frac{\mathbb{M}_D}{\mathbb{M}_R}$. In summary, the light neutrinos must mix LH flavor states only, while the heavy states mix the remaining RH fields. 
The matrix $\mathbb{M}_R$ is extracted from the part 
\begin{eqnarray}
\mathcal{L} &\supset& - \sum_{\alpha, \beta = 1,3} \frac{M_{\alpha \beta}}{2} \left[\overline{(\nu_{R\alpha})^c} \nu_{R \beta} + \overline{\nu_{R\alpha}} (\nu_{R \beta})^c \right]
\\ 
& &  - \sum_{\alpha = 1,3} Y_{\alpha 2} \left[\overline{(\nu_{R\alpha})^c} \nu_{R 2} + \overline{(\nu_{R2})^c} \nu_{R \alpha} \right] s  + Y_{\alpha 2}^* \left[\overline{\nu_{R2}} (\nu_{R \alpha})^c + \overline{\nu_{R\alpha}} (\nu_{R 2})^c \right] s^*
\end{eqnarray}
which can be presented in the basis
\begin{equation}
\mathcal{L} \supset 
\begin{pmatrix}
\overline{\nu}_{R1} & \overline{\nu}_{R2} & \overline{\nu}_{R3}
\end{pmatrix} 
\begin{pmatrix}
* & * & * \\ * & 0 & * \\ * & * & *
\end{pmatrix}
\begin{pmatrix}
{\nu}^c_{R1} \\ {\nu}^c_{R2} \\ {\nu}^c_{R3}
\end{pmatrix} + h.c.
\end{equation}
The mass eigenstates are denoted as $\chi_{R \alpha} = (\mathbb{V}_R)_{\alpha \beta} \nu_{R \beta}$. Once the rotation occurs among $\nu_R$'s, the fields $\chi$ are the degrees of freedom present in the interactions. As for the LH neutrinos, the $\chi$ spinors are given by
\begin{equation}
\chi = \chi_R + \chi^c_R\,. 
\end{equation}

The case $\mathbb{M}_D \ll \mathbb{M}_R$ is valid both from the presence of small Yukawas as from the difference between the electroweak and the Majorana scales. In the first scenario, small Yukawas, which control the decay  of the new mass eigenstates into SM particles (apart from the suppressed $\mathbb{V}_{LR}$), will dictate how reliable it is to assume at least one generation of $\chi$ fields as a dark matter candidate. 

The last part concerns neutrino interactions and a block diagonal U is assumed to rotate the system to the mass basis, such that the heavy neutrinos mixes the RH fields only, and the block $V_{\nu R}$ is approximately unitary (see \cite{Branco:1999fs}). The universal elements are weighted by
\begin{equation}
g_R^I(\nu_R) = g_Z^I(\nu_R) = 0\, , 
\end{equation}
since $\nu_R$ are singlets under the SM gauge group. Thus, 
\begin{eqnarray}\label{Eq.GIc}
\mathcal{L}_{kin} \supset s_\theta g_X \biggl[ \overline{\nu}_{R} V^\dagger_{\nu R} \mathbb{X}^\nu V_{\nu R} \gamma^\mu {\nu}_{R} \biggr] Z_\mu 
- c_\theta g_X \biggl[ \overline{\nu}_{R} V^\dagger_{\nu R} \mathbb{X}^\nu V_{\nu R} \gamma^\mu {\nu}_{R} \biggr] X_\mu\, ,
\end{eqnarray}

\subsection{\texorpdfstring{$Z$ Couplings}{}}
\begin{equation}
\mathcal{L} \supset \frac{1}{2} \ \overline{f} \ \gamma_\mu (g_V^f + g_A^f \gamma^5) \ f \ Z^\mu,
\end{equation}

\begin{subequations}
	\begin{eqnarray}
	g_V^U &=&  g \frac{c_\theta}{c_\phi} \left(\frac{1}{2} - \frac{4}{3} s^2_\phi \right)
	+ s_\theta \kappa \frac{5}{6} + s_\theta g_X \mathbb{F}^U_{ii},
	\\
	g_A^U &=&  g \frac{c_\theta}{c_\phi} \left(- \frac{1}{2} \right)
	+ s_\theta \kappa \frac{1}{2} + s_\theta g_X \mathbb{F}^U_{ii},
	\\
	g_V^D &=&  g \frac{c_\theta}{c_\phi} \left(- \frac{1}{2} + \frac{2}{3} s^2_\phi \right)
	- s_\theta \kappa \frac{1}{6} + s_\theta g_X \mathbb{F}^D_{ii},
	\\
	g_A^D &=&  g \frac{c_\theta}{c_\phi} \left(\frac{1}{2}\right)
	- s_\theta \kappa \frac{1}{2} + s_\theta g_X \mathbb{F}^D_{ii},
	\\
	g_V^l &=&  g \frac{c_\theta}{c_\phi} \left(- \frac{1}{2} + 2 s^2_\phi \right)
	- s_\theta \kappa \frac{3}{2} + s_\theta g_X \mathbb{F}^l_{ii},
	\\
	g_A^l &=& g \frac{c_\theta}{c_\phi} \left(\frac{1}{2}\right)
	- s_\theta \kappa \frac{1}{2} + s_\theta g_X \mathbb{F}^l_{ii}
	\\
	g_V^\nu &=&  - g_A^\nu = g \frac{c_\theta}{c_\phi}  \left(\frac{1}{2}\right)
	- s_\theta \kappa ,
	\\
	g_V^\chi &=& g_A^\chi = - s_\theta g_X
	\,.
	\end{eqnarray}
\end{subequations}
where the index $i$ denotes the fermion generation.

\subsection{\texorpdfstring{$K^0 - \overline{K^0}$}{}}
\begin{figure}[tbp]
	\centering 
	\includegraphics[width=.45\textwidth]{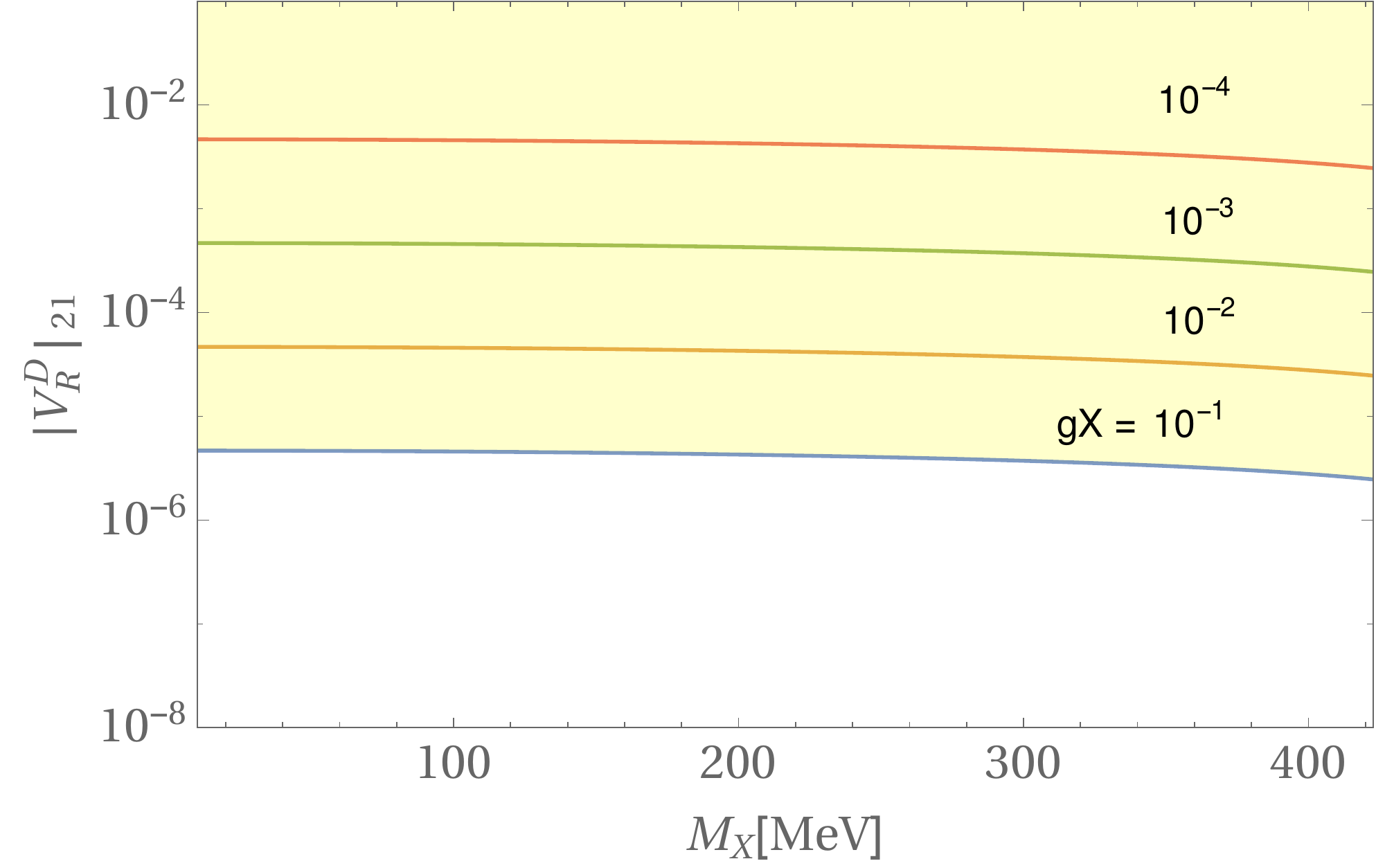}
	\caption{\label{fig:k0} The flavor changing matrix element $|V_d|_{21}$. The region above the lines is constrained according to \cite{Blum:2009sk}.}
\end{figure}

The contribution to the mass difference of the $K_{L,S}$, given by  $\Delta m_K$ in the $U(1)_X$ model corresponds to the operator 
\begin{equation}
\mathcal{H}_X = C(q^2) \ \overline{d} \gamma_\mu P_R s \ \overline{d} \gamma^\mu P_R s\,.
\end{equation}
Here the Wilson Coefficient comprises the $X^\mu$ propagator and the fermion couplings, i.e.
\begin{equation}
C(q^2) = - c_\theta^2 g_X^2 |\mathbb{F}^D_{ds}|^2 \frac{1}{q^2 - m_X^2} 
\end{equation}
Following the results in \cite{Blum:2009sk}, a bound can be imposed on $C(m^2_K)$ by claiming\footnote{Note: The same bound for LH currents has been considered once \begin{equation}
	\langle M^0 | (\overline{a} \gamma^\mu P_L q)(\overline{a} \gamma_\mu P_L q) |\overline{M^0} \rangle = \langle M^0 | (\overline{a} \gamma^\mu P_R q)(\overline{a} \gamma_\mu P_R q) |\overline{M^0} \rangle = \frac{f_M^2}{3} m_M
	\end{equation}}
\begin{equation}\label{Eq.KKbar}
|C(m^2_K)| \leq \frac{8.8 \times 10^{-19}}{[MeV]^2}
\end{equation}
In the region of the parameter space where $c_\theta \sim 1$, since $X_D = 1$, it can be converted into
\begin{equation}
g_X^2 
\frac{|V^d_{21}|^2}{m^2_K - m^2_X}
\leq \frac{8.8 \times 10^{-19}}{[MeV]^2}  
\end{equation}
where it has been taken $|V^d_{32}| \rightarrow 0$ and neglected quartic terms in $|V^d_{21}|$. In  Fig.\ref{fig:k0} the bounds to four values of the coupling $g_X$ are presented.

\acknowledgments
S.F.  acknowledges support of the Slovenian Research Agency under the core funding grant P1-0035.  F.C.C. would like to thank Clara H. Feliu, prof. Gudrun Hiller and prof. Emmanuel A. Paschos for useful discussions. F.C.C acknowledges support from the BMBF grant ``Verbundprojekt 05H2015:  Quark-Flavor-Physik am LHC (BMBF-FSP 105), Flavorsignaturen in Theorie und Experiment - LHCb: Run 2 and Upgrade'' and from the Technische Universit\"at Dortmund, Department of Physics.










\providecommand{\href}[2]{#2}\begingroup\raggedright\endgroup

\end{document}